\documentclass[a4paper,12pt]{article}
\usepackage{jheppub,esint,shuffle,psfrag}
\usepackage{bbm}
\usepackage{tikz}
\usetikzlibrary{arrows.meta,decorations.markings}

\usepackage{esint} 
\usepackage{breqn}
\usepackage{bm}

\def \tr {\mathop{\rm tr}\nolimits}
\def \Im {\mathop{\rm Im}\nolimits}
\def \Re {\mathop{\rm Re}\nolimits}

\newcommand\lr[1]{{\left({#1}\right)}}
\newcommand \widebar [1] {\overline{#1}}

\newcommand \vev [1] {\langle{#1}\rangle}

\newcommand \ket [1] {|{#1}\rangle}
\newcommand \bra [1] {\langle {#1}|}
\newcommand\re[1]{(\ref{#1})}
\def \qqquad {\qquad\quad}
\def \qqqquad {\qquad\qquad}

\newcommand{\rchi}{\mathord{\vcenter{\hbox{\raise 0.5ex\hbox{$\chi$}}}}}

\newcommand{\Wr}{\mathcal W}

\def\numberbysection{\@addtoreset{equation}{section}
                     \def\theequation{\thesection.\arabic{equation}}}
 
\begin{document}

\vspace*{0cm }

\author{Bercel Boldis$^{a,b}$, Nikolay Gromov$^{c}$, Gregory P. Korchemsky$^{d}$  and  Alessandro Testa$^{d}$  }
\affiliation{
$\null$
$^a$Department of Theoretical Physics, Institute of Physics, Budapest University of Technology and Economics M\H{u}egyetem rkp. 3, 1111 Budapest, Hungary
\\
$\null$ 
$^b$HUN-REN Wigner Research Centre for Physics, Konkoly-Thege Miklos ut 29-33, 1121 Budapest, Hungary
\\
$\null$ 
$^c$Mathematics Department, King’s College London, The Strand, London WC2R 2LS, UK
\\		
$\null$
$^d${Institut de Physique Th\'eorique\footnote{Unit\'e Mixte de Recherche 3681 du CNRS}, Universit\'e Paris Saclay, CNRS,  91191 Gif-sur-Yvette, France}   
}
\title{Localization identities for spectral determinants:
\\
two-dimensional QCD and Ising field theory}
\abstract{
Spectral determinants encode the meson spectra of planar two-dimensional QCD and of Ising field theory in the two-particle Bethe--Salpeter approximation. We derive localization identities for these  determinants which express their logarithmic derivatives with respect to the spectral parameter in terms of Baxter $Q-$functions evaluated at special points, proving previously conjectured relations. Our derivation uses finite-rank commutation relations and virial identities of the underlying integral spectral operators. We identify a sufficient rational closure condition on the symbol of these operators that extends localization to a broader class of spectral problems and determines the evaluation points. Applications include generalized two-dimensional QCD with a quartic gauge-field potential and symbols interpolating between the QCD and Ising cases. We also give a general construction of the spectral operators and establish conditions for the finite-rank reduction on which localization relies.
}
 
\maketitle
 
\section{Introduction}\label{sec:introduction}
Computing physical observables at finite coupling remains an open
problem in generic interacting quantum field theories. While perturbative and semiclassical expansions provide systematic approximations in specific limits, extending these results beyond their regimes of validity requires a deeper understanding of the underlying dynamics.
Recent developments have shown that
various observables in theories in $d=2,3,4$ dimensions admit exact
representations in terms of Fredholm determinants  
\begin{align}\label{intro:Fredholm}
D (\lambda)=\det(1-\lambda\bm K)=1+\sum_{n=1}^{\infty}\frac{(-\lambda)^n}{n!}
\int_{-\infty}^\infty dx_1\dots dx_n\,
\det \bigl[K(x_i,x_j)\bigr]_{i,j=1}^{n}\,,
\end{align}
where $\bm K$ is an integral operator acting on the real axis and its kernel $K(x,y)$ is a sufficiently regular function.
For a trace-class operator,  $D(\lambda)$ is an entire function of the spectral parameter, and  $D(\lambda)=\prod_n(1-\lambda\mu_n)$, where $\mu_n$ denoting the eigenvalues of $\bm K$.~\footnote{For the spectral problems arising in the two-dimensional models considered here, the relevant operators are Hilbert–Schmidt but not trace class. Therefore, their Fredholm determinants require regularization. \label{foot1}}
The form of the operator $\bm K$ is fixed by choosing an observable and its kernel encodes the dependence on the coupling and other physical parameters. Additional algebraic and analytic properties of these operators can simplify the evaluation of their
determinants, making these representations a practical tool to compute observables at finite coupling.

In two dimensions, examples include correlation functions of integrable
models~\cite{IIKS,Korepin:1993kvr} and the meson mass spectra of planar
QCD~\cite{Fateev:2009jf,Litvinov:2024riz} and of Ising field theory
in the two-particle Bethe--Salpeter
approximation~\cite{Litvinov:2025geb}. In three dimensions, Fredholm
determinants enter the Fermi--gas formulation \cite{Marino:2011eh,Hatsuda:2012hm,Boldis:2026zqu} of the partition function
and supersymmetric Wilson loops in ABJM
theory~\cite{Marino:2011eh,Klemm:2012ii,Hatsuda:2013yua,Boldis:2025yll}.

In four dimensions,
determinants of truncated Bessel operators govern correlation functions
of heavy half-BPS operators in planar $\mathcal N=4$ super Yang--Mills
theory, as well as free-energy corrections and correlators of chiral
primary operators in certain $\mathcal N=2$ superconformal theories
(see Ref.~\cite{Beccaria:2022ypy} and references therein).
These representations encode the exact dependence on the parameters of
the underlying theories. Their evaluation therefore gives access to
nonperturbative regimes, including exponentially suppressed
contributions that are invisible in a perturbative expansion.

Determinant representations provide systematic routes to the computation
of physical observables by bringing methods from spectral theory and
integrable systems to bear on quantum field theory. Fredholm
determinants have a long history in the study of integrable models,
where determinants of suitable kernels can be identified with tau
functions of integrable hierarchies or isomonodromic deformation
systems~\cite{IIKS,Korepin:1993kvr,Tracy:1993xj,Tracy:1993ah}. Their appearance across quantum field
theories in different dimensions thus hints at hidden integrable
structures governing the corresponding observables. A key feature of
these determinants is their ability to organize an infinite set of
spectral data into a single function. In two-dimensional QCD, for
example, the zeros of the spectral determinant determine the meson
masses. In the Fermi--gas description of ABJM theory,  the grand canonical partition function is described by a Fredholm determinant, whose expansion
in the fugacity generates the canonical partition functions at
different ranks of the gauge group.

The operators entering the determinant representations \re{intro:Fredholm} discussed above have different kernels, but they all belong  to the class of integrable operators introduced by Its, Izergin, Korepin and Slavnov~\cite{IIKS} in the study of correlation functions
in two-dimensional integrable models. After a  suitable change of variables, these kernels
can be written in the form
\begin{align}\label{intro:integrable}
K(x,y)=\frac{f_+(x)f_-(y)-f_-(x)f_+(y)}{x-y}\,,
\end{align}
where  $f_\pm(x)$ depend on the observable.
A distinguishing property of integrable operators is that the resolvent operator
$\bm K(1-\lambda\bm K)^{-1}$ retains the same structure, with the functions $f_\pm(x)$  
replaced by auxiliary functions $Q_\pm(x|\lambda)$. This property is the starting point of the Tracy--Widom approach~\cite{Tracy:1993xj,Tracy:1993ah,Tracy:1995ax}, which expresses logarithmic derivatives of Fredholm determinants as integrals of bilinear combinations of the functions $Q_\pm(x|\lambda)$ and their derivatives.
 
In general, the functions  $Q_\pm(x|\lambda)$ satisfy linear integro-differential equations determined by \re{intro:integrable}. It turns out that the kernels arising in QCD${}_2$, the Ising integral problem and
ABJM theory~\cite{Fateev:2009jf,Litvinov:2025geb,Boldis:2025yll}, in addition to being integrable,  have the following remarkable property: by analytically continuing  $Q_\pm(x|\lambda)$ to complex $x$, and using analyticity properties of the kernel, one can relate the values of $Q_\pm(x|\lambda)$ at  arguments shifted along the imaginary axis.
After a suitable normalization, the resulting relation takes the form of Baxter's TQ--equation \cite{Baxter:1972hz,Baxter1982}, 
\begin{align}\label{TQ}
\Delta_+(x)Q(x+2i|\lambda)
+\Delta_-(x)Q(x-2i|\lambda)
=T(x|\lambda)Q(x|\lambda)\,.
\end{align}
Here $Q$ denotes either of the two appropriately normalized auxiliary
functions; the dressing factors $\Delta_\pm$ and
the coefficient function $T$ are fixed by the integral problem. The appearance
of \re{TQ} is significant. In integrable models, Baxter's equation
relates transfer-matrix eigenvalues to auxiliary
$Q$-functions~\cite{Baxter:1972hz,Baxter1982}, and its emergence in Fredholm-determinant
calculations therefore suggests additional integrable structures in
the underlying field-theory observables.
 
The Baxter equation \re{TQ} alone does not simplify the calculation of  the determinant $D(\lambda)$. As mentioned above, the derivative $\partial_\lambda \log D(\lambda)$ is given by integrals involving bilinear combinations of the functions $Q_\pm(x|\lambda)$, schematically, (see, e.g. \cite{Tracy:1995ax})
\begin{align}\label{intro:dlogD}
 \partial_\lambda \log  D(\lambda) = \int_{-\infty}^\infty dx\,\rchi (x) \Big[Q'_+(x|\lambda) Q_-(x|\lambda)-Q'_-(x|\lambda) Q_+(x|\lambda)\Big]\,,
\end{align}
where $\rchi(x)$ is the so-called symbol function (see \re{chi-QCD} and \re{chi-Ising} below)   and the prime denotes the partial derivative with respect to $x$. Evaluating  \re{intro:dlogD}  requires detailed knowledge
of the $Q$-functions along the integration contour, including their
analyticity, growth at infinity  and normalization conditions.~\footnote{In general, the integral in \re{intro:dlogD} diverges at infinity and requires regularization (see footnote \ref{foot1}).} Since the 
Baxter equations  \re{TQ} generally have no known closed-form solutions,
the integral representation \re{intro:dlogD} by itself does not provide a readily
evaluable exact expression for the determinant.  
 
Our main goal is to prove that, in planar QCD${}_2$ and the Ising
field theory, the integrals entering the logarithmic derivatives
of the spectral determinants can be reduced to values of the Baxter
$Q$-functions and their $x$-derivatives evaluated at special points.
Such a simplification was conjectured by Fateev, Lukyanov and
Zamolodchikov~\cite{Fateev:2009jf} in their study of the 't~Hooft equation,
which determines the meson masses in two-dimensional QCD in the limit
$N_c\to\infty$ with the 't~Hooft coupling $g^2=g_{_{\rm YM}}^2N_c$ held fixed. For the special
quark masses $m_1^2=m_2^2=g^2/\pi$, their conjecture implies 
\begin{align}\label{intro:FLZ}
\partial_\lambda\log D(\lambda)
=2i\,\partial_x\log \Big(Q_+(x|\lambda)Q_-(x|\lambda)\Big)\Big|_{x=i} \ ,
\end{align}
with the zeros of $D(\lambda)$ giving the meson spectrum. Notice that the point $x=i$ lies away from the real integration contour in \re{intro:dlogD}. In this sense, the conjecture \re{intro:FLZ} ``localizes'' the integral: it replaces integration along the contour by the values of the $Q$-functions and their derivatives at a single unphysical point. Another identity conjectured in~\cite{Fateev:2009jf}
expresses the ratios of $Q$-functions evaluated at special points  through the ratio of even- and odd-sector determinants.

Subsequent work showed that this phenomenon persists beyond the special
mass choice. The relation \re{intro:FLZ} was generalized in~\cite{Ambrosino:2023dik,Litvinov:2024riz} to arbitrary equal quark masses $m_1=m_2$ and later extended in~\cite{Artemev:2025cev} to unequal masses $m_1\ne m_2$, together with the corresponding determinant-ratio identities. In this case, the relations become more complicated and involve suitable linear combinations of Baxter $Q-$functions. These
extensions substantially enlarged the range of accessible meson spectral data, while the localization identities remained conjectural.

An analogous structure was found in Ising field theory~\cite{Litvinov:2025geb}. In the two-particle approximation,
mesons are bound states described by the Bethe--Salpeter equation~\cite{Fonseca:2001dc,Fonseca:2006au}.
Its spectral problem also admits a Fredholm-determinant formulation, with an operator different from that of QCD${}_2$. The conjectured
localization relations involve Baxter functions evaluated at $x=2i$ and provide the Ising counterparts of \re{intro:FLZ}.~\footnote{Here one must distinguish the determinants of the common integrable kernel from the physical Bethe--Salpeter determinants, which are related
to them by a finite-rank modification.}

The localization identity \re{intro:FLZ} and its generalizations substantially simplify the evaluation of
the spectral determinants by reducing it to the analysis of Baxter's TQ--equations. The small-$\lambda$ expansions of $D(\lambda)$ generate spectral sums
that constrain the low-lying meson masses, while the large-$\lambda$
asymptotics yield quantization conditions and systematic expansions
for highly excited states. This strategy has been applied successfully
to planar QCD${}_2$ and to Ising field theory within the two-particle
Bethe--Salpeter approximation~\cite{Fateev:2009jf,Litvinov:2024riz,Artemev:2025cev,Litvinov:2025geb}.
Despite these advances and extensive analytic and numerical checks,
the localization identities have so far  lacked a
proof. In the present work, we establish these identities and identify
the algebraic and analytic properties of the integral operators
responsible for their simplification.

The paper is organized as follows. In Section~\ref{sect2}, we formulate the meson spectral problems in
terms of Fredholm determinants and introduce the underlying integral
operators. In Section~\ref{sect3}, we derive a virial
identity and use it to express spectral sums in terms of resolvent
matrix elements. In Section~\ref{sect4}, we
introduce the Baxter $Q$-functions and prove localization identities
for the logarithmic derivatives and the ratio of the sector
determinants. In Section~\ref{sec:rational-symbols}, we extend the
identities to symbols satisfying a rational closure condition and
apply the resulting formulas to a quartic gauge-field potential and
to symbols interpolating between the QCD and Ising cases.
In Section~\ref{sect6}, we generalize the integral kernels and
determine when the finite-rank virial reduction persists.
Concluding remarks
and directions for future work are given in Section~\ref{sect7}. Technical details are
collected in two appendices.
 
\section{Spectral determinants}\label{sect2} 

We begin by reviewing the meson bound-state equations in planar QCD${}_2$ and Ising field theory. Their mass spectra define spectral determinants through regularized infinite products over the bound-state levels. We then express these products as regularized Fredholm determinants of the corresponding integral operators, establishing the operator framework for the identities discussed in the Introduction.

\subsection{From meson bound states to spectral determinants}

In both theories, the mesons are described by a Bethe--Salpeter equation for a two-particle bound state. Its eigenvalues give the meson
masses, while its eigenfunctions describe the meson wave functions. In planar QCD${}_2$ this equation becomes the 't Hooft equation~\cite{tHooft:1974pnl} for bound states of quarks. In the ordered phase of Ising field theory, a magnetic field lifts the
degeneracy of the two vacua and confines a kink and an antikink into a
meson. The corresponding Bethe--Salpeter equation describes this bound
state in the two-particle approximation~\cite{Fonseca:2001dc,Fonseca:2006au}. In both cases the bound-state equation has a discrete spectrum of
masses, conveniently parametrized by dimensionless eigenvalues: $M_n^2=2\pi g^2\lambda_n$ in planar QCD${}_2$ and $M_n^2=4\pi f_0\lambda_n$ in Ising field theory. Here $g^2=g_{\rm YM}^2N_c$ is the 't Hooft coupling and $f_0$ is the
string tension in the Ising field theory. 

A spectral determinant combines the entire spectrum of the mesons  with masses  $M^2_n\sim \lambda_n$ into a single function of a complex spectral parameter $\lambda$
\begin{align}\label{D-def}
D(\lambda)= e^{\lambda d}\prod_{n\geq0}\lr{1-{\lambda\over\lambda_n}}\,e^{\lambda/\lambda_n}\,.
\end{align}
Finding meson masses becomes equivalent to finding zeros of this function.  
Expanding the logarithm of the spectral determinant near $\lambda=0$ produces sums of inverse
powers of the eigenvalues. At large $n$, the squared masses of highly excited states grow linearly, $\lambda_n\sim n/2$, and, as a consequence, the sum of their inverse eigenvalues has a logarithmic divergence. 
The additional exponential factor inside the product on the right-hand side of \re{D-def} is introduced to make the product convergent. The parameter $d$ in the prefactor in \re{D-def} yields a regularized sum of the inverse eigenvalues. Its value depends on the regulation scheme.

For generic quark masses, no closed-form solution of the 't Hooft equation is known, and the meson masses are usually computed numerically.\footnote{An efficient numerical approach is the Nystr\"om method~\cite{Nystrom:1930}, which replaces the integral operator by its quadrature discretization. Applied to Fredholm determinants, it converges exponentially fast in the number of quadrature nodes for kernels analytic in a neighborhood of the integration domain~\cite{Bornemann:2010}.} The advantage of the spectral determinants is that they can be analyzed using analytic tools.
These tools yield spectral sums and systematic asymptotic expansions with coefficients that retain the dependence on the quark
masses, without requiring a numerical solution for each bound-state
wave function~\cite{Fateev:2009jf,Litvinov:2024riz,Artemev:2025cev}.
The same strategy applies to the spectral problem in the Ising field theory~\cite{Litvinov:2025geb}.  

As shown in \cite{Fateev:2009jf,Artemev:2025cev,Litvinov:2025geb}, an integral transformation of the meson wave functions casts the Bethe--Salpeter equation for the meson spectrum in both theories into the form of a homogeneous Fredholm integral equation
\begin{align}\label{BS}
{1\over\rchi(x)}\psi_n(x)=\int_{-\infty}^\infty dy \, \Big[\lambda_n K(x-y) +{i\beta\over\pi} H(x-y)\Big]\psi_n(y)  \,.
\end{align}
It involves two convolution kernels defined as
\begin{align}\notag \label{HK}
{}&K(x-y) ={\pi(x-y)\over 2\sinh({\pi\over 2}(x-y))}\ ,
\\
{}&H(x-y)= {1\over \sinh({\pi\over 2}(x-y))}\,.
\end{align}
The integral involving $H(x-y)$ is understood using the principal value prescription,
whereas $K(x-y)$ has a regular diagonal limit, $K(0)=1$. In addition, the equation \re{BS} depends on the parameter $\beta$ and the function $\rchi(x)$, conventionally called the symbol. Their explicit expressions are different in the two theories and are given by
\begin{itemize}
\item Two-dimensional QCD with two flavours:
\begin{align}\label{chi-QCD}
{}&\rchi_{_{\rm QCD}}(x) =\left[ {{{2\alpha \over\pi}+x\coth\lr{\pi x\over 2}}}\right]^{-1},\qqqquad \beta={\pi\over 2g^2}(m_2^2-m_1^2)\,,
\end{align}
where $\alpha= \pi (m_1^2+m_2^2)/(2g^2)-1$, $g^2$  is the 't Hooft coupling constant and $m_i$ are the masses of the quark and the antiquark.
\item Ising field theory \footnote{The symbol (\ref{chi-Ising}) also appears in planar QCD$_2$ with scalar quarks \cite{Shei:1977ci}. The corresponding 't Hooft equation (\ref{BS}) has a non-vanishing $\beta$-term. The localization identities for the spectral determinants in this case were conjectured in \cite{Meshcheriakov:2025lsx}.}:
\begin{align}\label{chi-Ising}
{}&\hspace*{-25mm}\rchi_{_{\rm Ising}}(x) = \left[{{{2\alpha\over\pi}+x\tanh\lr{\pi x\over 2}}}\right]^{-1},\qqqquad   \beta=0\,,
\end{align}
where $\alpha= \pi m^2/(2f_0)$, $m$ is the mass of the kinks and $f_0$ is the string tension.
\end{itemize}
The two functions $\rchi_{_{\rm QCD}}(x)$ and $\rchi_{_{\rm Ising}}(x)$ are related to each other by replacing $\coth$ with $\tanh$. This allows us to study the two spectral problems within the same operator framework.  
  
 In the following sections, we focus on two-dimensional QCD and use
the symbol \re{chi-QCD}. The Ising case will be treated separately in Section~\ref{sec:applications}.
 
\subsection{Integral operators and their properties}\label{sec:KH-properties}

The convolution kernels in \re{HK} define operators on the real line
\begin{align}\label{KH-action}\notag
 {}& (\bm Kf)(x)=\int_{-\infty}^\infty K(x-y)f(y)\,dy\,,
 \\
 {}& (\bm Hf)(x)= \fint_{-\infty}^\infty H(x-y)f(y)\,dy\, ,
\end{align}
where the integral in the second relation is understood using the principal value prescription. It is convenient to Fourier transform the kernels \re{HK}  
\begin{equation}\label{KH-Fourier-kernels}
 \begin{aligned}
 K(x-y)&=\int_{-\infty}^\infty\frac{dp}{2\pi}\,
                 e^{ip(x-y)}\frac{\pi}{\cosh^2p},\\
 H(x-y)&=\int_{-\infty}^\infty\frac{dp}{2\pi}\,
                 e^{ip(x-y)}(-2i\tanh p),
 \end{aligned}
\end{equation}
and define the same operators in the momentum representation. 

We employ quantum-mechanical conventions and introduce the coordinate and momentum operators
\begin{equation}\label{KH-Fourier-convention}
 \vev{x|p}=\frac{e^{ipx}}{\sqrt{2\pi}}\,,\qqqquad
 \bm p=-i\partial_x\,,\qqqquad [\bm x,\bm p]=i.
\end{equation}
In this convention, both operators in \re{KH-action} are functions of the momentum operator
\begin{align}\label{KH-momentum}
 \bm K=\frac{\pi}{\cosh^2\bm p},\qqqquad
 \bm H=-2i\tanh\bm p.
\end{align}
In terms of these operators the Bethe--Salpeter equation \re{BS} takes the form
\begin{align}\label{eig}
\bm\rho(\lambda_n,\beta) \ket{\psi_n}=0 \, ,
\end{align}
where the notation was introduced for the spectral operator
\begin{align}\label{rho}
\bm\rho(\lambda,\beta)=\bm\rchi^{-1} -\lambda \bm K -{i\beta\over\pi} \bm H\,.
\end{align}
Here  $\bm\rchi^{-1}$ denotes multiplication by the function $1/\rchi(x)$. 
As follows from \re{chi-QCD} and \re{chi-Ising},  in both theories this function is  real, positive and regular on the
real axis.
 
\paragraph{Adjoint properties and spectral bounds.}  It follows from \re{HK} that the kernel $K(x-y)$ is real and even, whereas $H(x-y)$ is real and odd.
Consequently, the operator $\bm K$ is self-adjoint and $\bm H$ is skew-adjoint. Combined with \re{rho}, this leads to
\begin{align}
 \bm K^\dagger=\bm K\,,\qqqquad \bm H^\dagger=-\bm H\,,\qqqquad 
 \bm\rho^\dagger(\lambda,\beta)=\bm\rho(\lambda,\beta)\ ,
\end{align}
where $\lambda$ and $\beta$ are real. The last relation ensures that the eigenvalues $\lambda_n$ in \re{eig} are real.

Furthermore, it follows from  the definition of the operators \re{KH-momentum} that for the eigenstates normalized as $\vev{\psi_n|\psi_n}=1$
\begin{align}\label{ineq}
0< \vev{\psi_n|\bm K|\psi_n}\le \pi\,,\qqquad \vev{\psi_n|\bm\rchi^{-1}|\psi_n}\ge \rchi_{\rm max}^{-1}\,,\qqquad
|\vev{\psi_n|i\bm H|\psi_n}|\le 2\ ,
\end{align}
where $\rchi_{\rm max}=\sup_{x\in\mathbb R}\rchi(x)$ is the maximal value of the symbol function on the real axis. Combining these relations with 
\re{eig} we find that the eigenvalues satisfy
\begin{align}\label{bound1}
\lambda_n = {\vev{\psi_n|\bm\rchi^{-1}|\psi_n}-{\beta\over\pi} \vev{\psi_n|i\bm H|\psi_n}\over \vev{\psi_n|\bm K|\psi_n}}
\ge {1\over \pi}\lr{{1\over\rchi_{\rm max}}-{2|\beta|\over\pi}} \,.
\end{align}
For the symbol functions \re{chi-QCD} and \re{chi-Ising}, this yields the lower bounds for the eigenvalues in QCD${}_2$ and the Ising field theory 
\begin{align}\label{bound2}
\lambda_n^{\rm {}_{QCD}}\ge{2\over\pi^2}(\alpha+1-|\beta|)\,,\qqqquad
\lambda^{\rm Ising}_n\ge{2\over\pi^2}\alpha\,.
\end{align}
Replacing $\alpha$ and $\beta$ with their values \re{chi-QCD} and \re{chi-Ising}, it is easy to see that 
both bounds are non-negative, ensuring that the squared meson masses $M_n^2$ are non-negative in both theories.
 
\paragraph{Symmetries.}  Let us introduce the parity transformation $(\bm Pf)(x)=f(-x)$ and complex
conjugation $(\bm Cf)(x)=\bar f(x)$. The integral operators \re{KH-momentum} satisfy
\begin{align}\label{KH-parity} 
 \bm P\bm K\bm P=\bm K,\qqquad \bm P\bm H\bm P=-\bm H,
\qqquad \bm C\bm K\bm C=\bm K,\qqquad \bm C\bm H\bm C=\bm H.
\end{align}
Accordingly, $\bm K$ preserves parity, while $\bm H$ exchanges even and odd functions. The combination $i\beta\bm H$ is self-adjoint for
real $\beta$ and invariant under the combined action $\bm C\bm P$. As a consequence, for an even symbol function, $\rchi(-x)=\rchi(x)$, the spectral operator \re{rho} satisfies
\begin{align}
 \bm P\bm\rho(\lambda,\beta) \bm P=\bm C\bm\rho(\lambda,\beta) \bm C=\bm\rho(\lambda,-\beta)\,.
\end{align}
It follows from this relation that for $\beta=0$ the eigenstates of $\bm\rho(\lambda,0)$ can be separated according to their parity. This happens in QCD${}_2$ for equal masses, $m_1=m_2$, and in the Ising field theory. In the latter case, fermionic statistics selects odd wave functions as physical mesons. The even solutions are useful auxiliary states. For $\beta\neq 0$, the eigenstates of $\bm\rho(\lambda,\beta)$ can be separated using the virial identity (see \re{vir-eq1} below) by
continuation from the odd and even sectors at $\beta=0$. 

\paragraph{Commutation relations.} Since both operators \re{KH-momentum} are functions of $\bm p$, they commute. In
addition, the relation $K(x-y)=\pi(x-y)H(x-y)/2$ and the identity
$1-\tanh^2p=1/\cosh^2p$ give
\begin{equation}\label{KH-algebra}
 [\bm K,\bm H]=0,\qqqquad
 \bm K=\frac\pi2[\bm x,\bm H]
       =\frac\pi4(\bm H^2+4).
\end{equation}
We use the identity
\begin{align} 
{}&\left[\tanh\Big({\pi\over 2}x\Big)-\tanh\Big({\pi\over 2}y\Big)\right]H(x-y)= {1\over \cosh({\pi\over 2}x)\cosh({\pi\over 2}y)} 
\end{align}
to obtain the operator relations
\begin{align}\label{H-com} 
{}&[\bm t,\bm H] = \ket{e_+}\bra{e_+}\,, \qqqquad 
 [\bm t,\bm K] ={\pi\over 2}\Big(\ket{e_-}\bra{e_+} - \ket{e_+}\bra{e_-}\Big)\, ,
\end{align}
where $\bm t=\tanh(\pi\bm x/2)$. 
The second relation follows from applying \re{KH-algebra}. 
Here  the operator $\bm t$ denotes multiplication by $\tanh(\pi x/2)$ and the notation was introduced for two reference states
\begin{align}\label{epm}
e_+(x) = {1\over \cosh({\pi\over 2}x)}\,,\qqqquad e_-(x) = {x\over \cosh({\pi\over 2}x)}\,.
\end{align}
The subscript refers to the parity of the states, $\bm P e_\pm(x) = \pm e_\pm(x)$.

Notice that $\bm H$ and $\bm K$ are infinite-rank operators, whereas their commutators \re{H-com}  are operators of rank $1$ and rank $2$, respectively. It is this property that makes the operator $\bm K$ integrable.  Indeed, up to a constant, the kernel of this operator defined in \re{HK} coincides with \re{intro:integrable} upon replacing $x$ with $t(x)=\tanh(\pi x/2)$ and $f_\pm(x)$ with $e_\pm(x)$.  

\paragraph{Reference states.} Fourier transforming \re{epm}, we can obtain operator representations of the reference states
\begin{align}
\ket{e_+}={2\over \cosh\bm p}\ket{0}\,,\qqqquad \ket{e_-}=2\bm x{1\over \cosh\bm p}\ket{0}=-2i{\sinh\bm p\over \cosh^2\bm p} \ket{0}\,,
\end{align}
where $\vev{x|0}=\delta(x)$. Combining these relations with \re{KH-momentum} and \re{KH-algebra}, we can determine the action of the operators $\bm H$ and $\bm K$ on the reference states
\begin{align}\label{c-rel}\notag
{}& \bm H \ket{e_+}=2\ket{e_-}\,, && \bm H  \ket{e_-}=(\bm x^2-1)\ket{e_+}\,, 
\\[2mm]
{}&\bm K \ket{e_+}={\pi\over 2}(\bm x^2+1)\ket{e_+} \,, && \bm K \ket{e_-}={\pi\over 6}\bm x(\bm x^2+1)\ket{e_+}  \,.
\end{align}
The same relations can also be obtained by convolving  the kernels \re{HK} with the functions \re{epm}.
 
\subsection{Connection to Fredholm determinants}

We first consider $\beta=0$. Writing $\psi_n(x)=\rchi(x) \Psi_n(x)$, the spectral equation \re{eig} takes the form 
\begin{equation}\label{compact-eigenproblem}
 \bm K\bm\rchi\ket{\Psi_n}=\mu_n\ket{\Psi_n}\,,
\end{equation}
where $ \mu_n=1/\lambda_n$. In  close analogy with \re{intro:Fredholm} we can define the spectral determinant as the formal Fredholm product
\begin{align}\label{D-naive}
\det(\bm\rho(\lambda,0)\bm\rchi)= \det(1-\lambda \bm K\bm\rchi)=\prod_n (1-\lambda \mu_n)\,,
\end{align}
but it does not converge. The term linear in $\lambda$ is proportional to the integral $\tr(\bm K\bm\rchi)=\int_{\mathbb{R}} dx \,\rchi(x)$, which diverges because $\rchi(x)\sim1/|x|$ at large $x$ for the symbol functions in \re{chi-QCD} and \re{chi-Ising}. The regularized spectral determinant can be defined as a normalized second Fredholm determinant~\footnote{For a Hilbert--Schmidt operator $\bm A$, this determinant is defined as ${\det}_2(1-\bm A)=\det\left[(1-\bm A)e^{\bm A}\right]$.}
\begin{equation}\label{spectral-Fredholm}
 D(\lambda)\Big|_{\beta=0}=e^{d\lambda}\, {\det}_2\left(1-\lambda \bm K\bm\rchi\right).
\end{equation}When written in terms of the eigenvalues, this expression matches \re{D-def}, with the factors $e^{\lambda/\lambda_n}$ providing precisely the subtractions required by $\det_2$.
 
For $\beta\neq 0$,  the spectral equation \re{eig}  takes the form 
\begin{align}\label{mu_n}
\left(1-{i\beta\over\pi}\bm H\bm\rchi\right)^{-1}\bm K\bm\rchi\ket{\Psi_n}=\mu_n\ket{\Psi_n}\,.
\end{align}
One can show using \re{ineq} that for $\alpha+1-|\beta|\ge 0$ the operator $1-{i\beta\over\pi}\bm H\bm\rchi$ has positive eigenvalues and its inverse operator is well defined. In close analogy with \re{spectral-Fredholm}, the regularized spectral determinant is defined as
\begin{align}\label{D-sp}
D(\lambda)=e^{d\lambda}\,{\det}_2\left[1-\Big(1-{i\beta\over\pi}\bm H\bm\rchi\Big)^{-1}\lambda \bm K\bm\rchi \right]\,.
\end{align}
As before, the spectral determinant \re{D-sp} admits the representation \re{D-def} in terms of the eigenvalues $\mu_n=1/\lambda_n$. For arbitrary $\beta$ the logarithmic derivative of the spectral determinants \re{spectral-Fredholm} and \re{D-sp} is given by
\begin{equation}\label{spectral-logderivative}
 \partial_\lambda\log D(\lambda)
 =d-\lambda\sum_{n\ge 0}\frac{\mu_n^2}{1-\lambda\mu_n},
\end{equation}
where the sum over the spectrum is convergent. The finite constant $d=\partial_\lambda\log D(0)$ yields a  regularized sum of inverse eigenvalues
\begin{align}\label{d}
d=c-\sum_{n\ge 0}\Big(\mu_n -{2\over n+1}\Big)\,.
\end{align}
At large $n$ the inverse eigenvalues behave as $\mu_n=2/(n+1)+ O(\log n/n^2)$ and the sum in \re{d} converges \cite{Fateev:2009jf,Artemev:2025cev}.

The value $d$ depends on the subtraction convention. A change of convention, parametrized by $c\to c+\delta$, induces the multiplicative ambiguity of the regularized spectral determinant
$D(\lambda)\to e^{\lambda\delta}D(\lambda)$. 
    
\section{Spectral problem}\label{sect3} 

The spectral determinant \re{D-def} is a regularized product over the
spectral values $\lambda_n=1/\mu_n$ defined by \re{mu_n}. In this
section, we use the algebraic properties of $\bm H$ and $\bm K$ to compute $D(\lambda)$ in terms of the expectation values of the inverse spectral operator $\bm\rho^{-1}(\lambda,\beta)$. 

It is convenient to introduce the operator 
\begin{align}\label{M}
\bm M=\bm\rchi^{-1}-{i\beta\over\pi}\bm H \ .
\end{align}
The equation \re{eig} can then be written as the generalized
eigenvalue problem 
\begin{align}\label{eig1}
\bm K \ket{\psi_n}= \mu_n \bm M \ket{\psi_n}\,.
\end{align}
The operators $\bm M$ and $\bm K$ are Hermitian for real $\beta$, but they do not commute, $[\bm M,\bm K]=[\bm\rchi^{-1},\bm K]\neq 0$.
In addition, the operator $\bm M$ is strictly positive. The eigenvalues 
$\lambda_n=1/\mu_n$ satisfy the bounds given in \re{bound1} and \re{bound2}. The
eigenstates can be chosen orthonormal in the $\bm M$-weighted inner
product,
\begin{align}\label{inner}
\vev{\psi_n|\bm M|\psi_m}=\delta_{nm}\,.
\end{align}
With this normalization, completeness gives the spectral expansion of the operator $\bm\rho(\lambda,\beta)=\bm M -\lambda \bm K$ and its inverse  
\begin{align} \label{dec}\notag
{}&\bm\rho (\lambda,\beta)=\sum_{n\ge 0}(1-\lambda\mu_n)  \bm M\ket{\psi_n}\bra{\psi_n}\bm M \ ,
\\
{}&\bm\rho^{-1}(\lambda,\beta)=\sum_{n\ge 0} {\ket{\psi_n}\bra{\psi_n}\over 1-\lambda\mu_n}\ .
\end{align}
The second expression holds away from the spectral values
$\lambda=\lambda_n$, where the inverse develops poles.

\subsection{Virial identity} 

Let $\bm q$ be an operator, not necessarily Hermitian, for which the following transformation is well defined 
\begin{align}\label{rho-eps}
\bm\rho_\epsilon(\lambda,\beta)=e^{-i\epsilon \bm q^\dagger} \bm\rho(\lambda,\beta)e^{i\epsilon \bm q} =  \bm\rho(\lambda,\beta)+ i \epsilon \left[\bm\rho(\lambda,\beta)\bm q- \bm q^\dagger \bm\rho(\lambda,\beta) \right] + O(\epsilon^2)\,.
\end{align}
For real $\epsilon$ and $\lambda$, this transformation preserves the hermiticity of $\bm\rho_\epsilon(\lambda,\beta)$. At $\lambda=\lambda_n$, the eigenvalue equation \re{eig1}
implies $\bm\rho(\lambda_n,\beta)\ket{\psi_n}=0$. Consequently, the first two terms in the small-$\epsilon$ expansion of the expectation value of \re{rho-eps} vanish,
$
\vev{\psi_n|\bm\rho_\epsilon(\lambda_n,\beta)|\psi_n}
=O(\epsilon^2)
$.
In particular, we obtain the virial identity
\begin{align}\label{vir}
\vev{\psi_n| \bm\rho(\lambda_n,\beta)\bm q
-\bm q^\dagger\bm\rho(\lambda_n,\beta) |\psi_n}=0\,.
\end{align}
As we show below, a suitable choice of $\bm q$ turns this identity into a powerful relation between
the eigenvalues $\mu_n$ and the corresponding eigenfunctions $\psi_n(x)$. 

Let us choose the operator $\bm q$ in \re{vir} to be
\begin{align}\label{Q=t}
\bm q=\bm t = \tanh\Big({\pi \bm x\over 2}\Big) \ .
\end{align}
This operator is Hermitian and odd under parity, $\bm P\bm t\bm P=-\bm t$, so it exchanges even and odd states.
Since $\bm t$ commutes with $\bm\rchi^{-1}$, the commutation relations \re{H-com} give 
\begin{align}\notag\label{aux1}
[\bm\rho(\lambda_n,\beta),\bm t]{}&=[\bm M -\lambda_n \bm K,\bm t]= {i\beta\over\pi} [\bm t,\bm H]+\lambda_n[\bm t,\bm K]
\\[1.2mm]
{}&={i\beta\over\pi}\ket{e_+}\bra{e_+}
+{\pi\lambda_n\over2}\lr{\ket{e_-}\bra{e_+}-\ket{e_+}\bra{e_-}} \ .
\end{align}
Taking its expectation value and applying \re{vir}, we find
\begin{align}\label{vir1}
{i\beta\over\pi} |\vev{e_+|\psi_n}|^2+{\pi\lambda_n\over2}\Big(\vev{\psi_n|e_-}\vev{e_+|\psi_n}-\vev{\psi_n|e_+}\vev{e_-|\psi_n}\Big)=0\,.
\end{align}
Notice that the operators $\bm M$ and $\bm K$ are invariant under the combined $\bm {CP}$ transformations \re{KH-parity}. Since $(\bm{CP})^2=1$, we can choose a basis of
eigenstates invariant under this anti-linear symmetry, with phases fixed
so that
$\widebar\psi_n(x)=\psi_n(-x)$. The reference functions $e_\pm(x)$
are real and have definite parity, which implies
\begin{align}\label{psi-e}
\vev{\psi_n|e_\pm}=\pm\vev{e_\pm|\psi_n}\,.
\end{align}
Thus, the matrix elements $\vev{\psi_n|e_+}$ and $\vev{\psi_n|e_-}$ take real and purely imaginary  values, respectively.
Substituting these relations into \re{vir1} yields
\begin{align}\label{vir-eq1}
\vev{e_+|\psi_n}\left[{i\beta\over\pi^2}\vev{e_+|\psi_n} - \lambda_n \vev{e_-|\psi_n}\right]=0\,.
\end{align}    
This equation holds for all eigenstates satisfying \re{eig1}. It gives two possible selection rules, according to
which factor vanishes.  

At $\beta=0$, the eigenfunctions can be chosen
real and of definite parity, $\psi_n(-x)=(-1)^n\psi_n(x)$.
Parity implies that the odd states $\psi_{2n+1}(x)$ have zero overlap with $e_+(x)$, while the even states $\psi_{2n}(x)$ have zero overlap with $e_-(x)$. 
Continuing these two spectral sequences away
from $\beta=0$, we label their states by odd and even indices and write
the selection rules as
\begin{align}\notag\label{sec}
{}&\vev{e_+|\psi_{2n+1}} =0\,,
\\[2mm]
{}&\vev{e_-|\psi_{2n}}
-{i\beta\over\pi^2}\mu_{2n}\vev{e_+|\psi_{2n}} =0\,.
\end{align}
For $\beta\ne0$, the states $\psi_{2n+1}(x)$ and $\psi_{2n}(x)$ no longer have definite parity. Their labels are assigned by continuation from the odd and even sectors at $\beta=0$, respectively.

Applying the relations \re{sec}, we can separate all solutions to the spectral problem \re{eig1} into two sectors parameterized by the eigenspectrum $\{\mu_{2n}, \psi_{2n}(x)\}$ and $\{\mu_{2n+1}, \psi_{2n+1}(x)\}$ (with $n\ge 0$), respectively. The eigenstates belonging to these sectors are orthogonal to each other with respect to the inner product \re{inner}. 

We can obtain another relation for the eigenspectrum by applying the virial identity \re{vir} with the operator $\bm q$ of the form 
\begin{align}\label{q-def}
\bm q= i \bm t  \bm H \,,
\end{align} 
where the operator $\bm t$ is defined in \re{Q=t}.
This operator is non-Hermitian, $\bm q^\dagger =i \bm H \, \bm t$. It is even under the parity $\bm P$ and odd under the combined $\bm {CP}$ transformations \re{KH-parity}. 
Going through the calculation we find
\begin{align}\notag\label{aux2}
 \bm\rho(\lambda_n,\beta)\,\bm t \, \bm H -\bm H \, \bm t\,\bm \rho(\lambda_n,\beta) {}&= [\bm t \bm\rchi^{-1},\bm H]- \lambda_n (\bm K \bm t \bm H - \bm H \bm t \bm K)
 \\\notag
 {}&={2\over\pi}\bm K +{2\alpha\over\pi} [\bm t,\bm H]-{\pi\lambda_n\over 4}  (4 [\bm t, \bm H]-\bm H [\bm t,\bm H] \bm H )
 \\
 {}&={2\over\pi}\bm K + {2\alpha\over\pi}\ket{e_+}\bra{e_+} - \pi \lambda_n\Big(\ket{e_+}\bra{e_+}+ \ket{e_-}\bra{e_-} \Big)\,.
\end{align}
Here in the first relation we used that the operators $\bm t$ and $\bm\rchi$ commute, applied \re{KH-algebra}
 and replaced the symbol function $\rchi(x)$ by its explicit expression \re{chi-QCD} in QCD${}_2$.
In the second relation, we used the commutation relations \re{H-com} together with \re{c-rel}. Notice that the right-hand side of \re{aux2} is independent of $\beta$.

Taking the expectation value of \re{aux2} and applying \re{vir}, we find
 \begin{align}\label{qcd}
 \mu_n^2 = {\pi^2\over 2}\Big(|\vev{e_+|\psi_n}|^2+|\vev{e_-|\psi_n}|^2\Big)- \alpha \mu_n|\vev{e_+|\psi_n}|^2\,.
\end{align}      
In the two sectors defined in \re{sec}, this relation can be simplified as      
\begin{align}\notag\label{mus}
{}&\mu_{2n+1}^2 ={\pi^2\over 2}\left|\vev{e_-|\psi_{2n+1}}\right|^2 \,,
\\
{}&\mu_{2n}^2 ={\pi^2\over 2} \left[1+ {\beta^2\over\pi^4}\mu_{2n}^2-{2\alpha\over \pi^2} \mu_{2n}\right]\left|\vev{e_+|\psi_{2n}}\right|^2 \,.
\end{align}       
The relations \re{sec} and \re{mus} play complementary roles: \re{sec} defines the two sectors of eigenstates, while \re{mus} expresses the eigenvalues in terms of the overlaps of the corresponding eigenstates with the reference states \re{epm}.
\subsection{Spectral sums}

The decomposition into the two spectral sectors induces a factorization of the regularized spectral determinant \re{D-def} into the product of the determinants defined in each sector
\begin{align}
D(\lambda) = D_+(\lambda) D_-(\lambda)\,.
\end{align}
Here $D_+(\lambda)$ and $D_-(\lambda)$ are given by the product over the eigenvalues $\mu_{2n}$ and $\mu_{2n+1}$, respectively. 
They satisfy the relations analogous to \re{spectral-logderivative}
\begin{align}\notag\label{Dpm}
{}&\partial_\lambda\log D_+(\lambda)=d_+-\lambda\,\sum_{n\ge 0}\frac{\mu_{2n}^2}{1-\lambda\mu_{2n}}\,,
\\
{}&\partial_\lambda\log D_-(\lambda)=d_--\lambda\,\sum_{n\ge 0}\frac{\mu_{2n+1}^2}{1-\lambda\mu_{2n+1}}\,,
\end{align}
where the constants $d_\pm$ are related to the constant \re{d} as $d=d_++d_-$. 

By applying the virial relations \re{mus}, we can express the convergent sums in \re{Dpm} by  the so-called potentials. These are defined as expectation values of the inverse spectral operator in the reference states \re{epm}, namely
\begin{align}\label{V++}\notag
{}&V_{++}(\lambda) = \vev{e_+ |\bm \rho^{-1}(\lambda,\beta)|e_+}=\sum_{n\ge 0} {|\vev{e_+|\psi_n}|^2\over 1-\lambda\mu_n}\,,
\\
{}&V_{--}(\lambda) = \vev{e_- |\bm \rho^{-1}(\lambda,\beta)|e_-}=\sum_{n\ge 0} {|\vev{e_-|\psi_n}|^2\over 1-\lambda\mu_n}\,.
\end{align}
The spectral representations on the right-hand side follow from \re{dec}. The dependence of the potentials on the mass parameter $\beta$ is tacitly assumed.
The selection rules \re{sec} imply that $V_{++}$ receives contributions only from the even-labelled sector, whereas $V_{--}$ generally receives
contributions from both sectors.
 
Consider first the odd-labelled sum entering the second relation in \re{Dpm}. Using the first relation in
\re{mus}, and subtracting the even-labelled contribution to $V_{--}(\lambda)$,
we obtain
\begin{align}\notag
\sum_{n\ge0}{\mu_{2n+1}^2\over1-\lambda\mu_{2n+1}}&={\pi^2\over2}\bigg[V_{--}(\lambda)-\sum_{n\ge0}{|\vev{e_-|\psi_{2n}}|^2\over1-\lambda\mu_{2n}}\bigg]
\\
&={\pi^2\over2}\bigg[V_{--}(\lambda)-{\beta^2\over\pi^4}\sum_{n\ge0}{\mu_n^2|\vev{e_+|\psi_n}|^2\over1-\lambda\mu_n}\bigg]\,.
\end{align}
In the second equality, we used the even-sector relation in \re{sec} and then extended the remaining sum to the full spectrum by taking into account that 
odd-labelled terms vanish.  The sum in the second relation can be evaluated by subtracting the first terms of
the Taylor expansion of $V_{++}(\lambda)$. 

In  this way, we arrive at 
\begin{align}\label{dD-}
\partial_\lambda\log D_-(\lambda)=d_--\lambda\,{\pi^2\over 2}\Big[V_{--}(\lambda) -{\beta^2\over\pi^4 \lambda^2}\Big( V_{++} (\lambda) -V_{++} (0)-\lambda V_{++} '(0)\Big)\Big],
\end{align}
where the prime denotes differentiation with respect to $\lambda$. The additional terms proportional to $V_{++} (0)$ and $V_{++} '(0)$ are needed to ensure that the right-hand side remains finite for $\lambda\to 0$.  Applying the second relation in \re{mus} to the even-labelled sum in \re{Dpm}
and performing the analogous Taylor subtractions, we obtain
\begin{align}\notag\label{dD+}
\partial_\lambda\log D_+(\lambda)=d_+-\lambda\,{\pi^2\over 2}\Big[ V_{++}(\lambda){}&-{2\alpha\over\pi^2\lambda}\Big(V_{++}(\lambda)-V_{++} (0)\Big)
\\
{}&+ {\beta^2\over\pi^4 \lambda^2}\Big( V_{++} (\lambda) -V_{++} (0)-\lambda V_{++} '(0)\Big)\Big].
\end{align}
Combined with the boundary conditions, $D_\pm(0)=1$, these relations allow us to compute the two spectral determinants through the
potentials $V_{++}(\lambda)$ and $V_{--}(\lambda)$, up to the constants $d_\pm$.

Combining \re{dD-} and \re{dD+}, the terms  proportional to $\beta^2$ cancel, yielding 
\begin{align}
\partial_\lambda\log D(\lambda)
=d-{\pi^2\lambda\over2}\Big[V_{++}(\lambda)+V_{--}(\lambda)\Big]
+\alpha\Big[V_{++}(\lambda)-V_{++}(0)\Big]\,.
\end{align}
The right-hand side retains its dependence on $\beta$ through the potentials, $V_{++}$ and $V_{--}$, and the normalization constant $d$.

\section{Localization identities} \label{sect4}

In this section, we introduce the Baxter functions $Q_\pm(x)$ and derive two localization identities for the sector determinants $D_\pm(\lambda)$. The first expresses the 
logarithmic derivatives \re{dD-} and \re{dD+} through logarithmic derivatives of the Baxter functions at $x=i$. The second expresses $D_-(\lambda)/D_+(\lambda)$ through their values at $x=0$.

\subsection{Baxter $Q$-functions}
\label{sec:BaxQCD}
For real $x$, the functions $Q_\pm(x)$ are defined as the solutions of the inhomogeneous
Fredholm equations
\begin{align}\label{Q-defs}
Q_\pm(x)=E_\pm(x)
+{1\over e_+(x)}\fint_{-\infty}^{\infty}dy\,
\left[\lambda K(x-y)+{i\beta\over\pi}H(x-y)\right]
 e_+(y)\rchi(y)Q_\pm(y)\,,
\end{align}
where $E_+(x)=1$ and $E_-(x)=x$. The kernels are given in \re{HK},
while $\rchi(x)$ and $e_\pm(x)$ are defined in \re{chi-QCD} and \re{epm}.~\footnote{Alternatively, one may choose the reference states $e_-(x)=1/\sinh(\pi x/2)$ and $e_+(x)=xe_-(x)$, as in Refs.~\cite{Fateev:2009jf,Litvinov:2024riz}. The corresponding Baxter functions satisfy the same TQ-equation (see \re{Q-TQ} below) and can be expressed as linear combinations of the functions $Q_\pm(x)$. The localization identities are independent of this choice. The reference states \re{epm} have the advantage of being regular at $x=0$.}
Only the term involving $H$ requires the principal-value prescription.
We suppress the dependence of $Q_\pm(x)$ on the spectral parameter $\lambda$ and the mass parameters $\alpha$ and
$\beta$ unless it is needed explicitly.

The operator form of \re{Q-defs} is
\begin{align}\notag\label{Q-op}
Q_\pm(x)
&={1\over e_+(x)\rchi(x)}
\bra{x}\bm\rho^{-1}(\lambda,\beta)\ket{e_\pm}
\\
&={1\over e_+(x)}
\bra{x}{1\over1-(\lambda\bm K+i\beta\bm H/\pi)\bm\rchi}
\ket{e_\pm}\,,
\end{align}
where $e_-(x)$ is given by \re{epm}. These relations hold away from the spectral values where the inverse
operator develops poles.
 
For $\beta=0$, the functions $Q_\pm(x)$ are analytic in the fundamental strip $-2<\Im x<2$. Indeed, the kernel $K(x-y)$ is regular at 
$x=y$ and its nearest poles lie at $x-y=\pm2i$. The relevant Baxter solutions grow at large $x$ inside the strip more slowly than any exponential.
Since $1/e_+(x)=\cosh(\pi x/2)$ vanishes for $x=\pm i$, it follows from \re{Q-defs} that 
\begin{align}\label{Q-b}
Q_+(\pm i) =1 \,,\qqqquad Q_-(\pm i) = \pm i\ .
\end{align} 

For $\beta\ne0$, due to the pole of $H(x-y)$ at $x=y$ we have to distinguish two different continuations of \re{Q-defs} to complex $x$ -- the direct integral continuation and the Baxter continuation. The direct integral continuation is obtained by replacing the argument $x$ in \re{Q-defs} by a complex variable $z$, while keeping
the integration contour and the solution in the integrand on the real axis
\begin{align}\label{Q-direct-continuation}
\widetilde Q_\pm(z)=E_\pm(z)
+{1\over e_+(z)}\int_{-\infty}^{\infty}dy\,
\left[\lambda K(z-y)+{i\beta\over\pi}H(z-y)\right]
e_+(y)\rchi(y)Q_\pm(y)\,.
\end{align}
The poles of the kernels in \re{Q-direct-continuation} produce cuts of $\widetilde Q_\pm(z)$
along the lines $\Im z=2n$ with $n\in\mathbb Z$. In particular, the cut
along the real axis comes from the pole
$H(z-y)=2/[\pi(z-y)]+O(z-y)$ at $z=y$. Applying the
Sokhotski--Plemelj formula to this pole gives the upper and lower
boundary values
\begin{align}\label{Q-boundary}
\widetilde Q_\sigma(x\pm i0)
=\Delta_\pm(x)Q_\sigma(x)\,,\qquad \sigma=(+,-)\,,
\end{align}
where the notation was introduced for the dressing factors
\begin{align}\label{Q-dressing-factors}
\Delta_+(x)=1+{2\beta\over\pi}\rchi(x)\,,\qqqquad
\Delta_-(x)=1-{2\beta\over\pi}\rchi(x)\,.
\end{align}
The subscripts in $\Delta_\pm(x)$ refer to the upper and lower boundary values. For $\beta\ne0$, the
direct boundary values $\widetilde Q_\sigma(x\pm i0)$ therefore differ from the principal-value
solutions $Q_\pm(x)$ on the real axis.

The Baxter continuation $Q_\pm(z)$ is defined by sewing together the two branches defined above and below the real axis
\begin{align}\label{Q-Baxter-continuation}
Q_\pm(z)=
\begin{cases}
\displaystyle {\widetilde Q_\pm(z)/
\Delta_+(z)},&\ \ 0<\Im z<2,
\\[4mm]
\displaystyle {\widetilde Q_\pm(z)/\Delta_-(z)},&-2<\Im z<0.
\end{cases}
\end{align}
Both branches have the same boundary value $Q_\pm(x)$ on the real axis  and
define a single meromorphic function in the fundamental strip.
Its possible poles occur at zeros of the denominators in
\re{Q-Baxter-continuation}, unless canceled by zeros of the numerator.
For equal masses, $\beta=0$, the dressing factors reduce to unity and
the Baxter continuation coincides with the direct integral continuation.
 
By the same argument as before, the function $\widetilde Q_\pm(z)$ defined in \re{Q-direct-continuation} satisfies the relations
\begin{align}\label{Q-special-raw}
\widetilde Q_+(\pm i)=1\,,\qqqquad
\widetilde Q_-(\pm i)=\pm i\, ,
\end{align}
which coincide with  \re{Q-b} for $\beta=0$.  For $\beta\neq 0$, we apply \re{Q-Baxter-continuation} to find
\begin{align}\label{Q-i}
Q_+(\pm i)={\alpha\over\alpha\pm\beta}\,,\qquad
Q_-(\pm i)=\pm i\,{\alpha\over\alpha\pm\beta}\,,
\end{align}
where we took into account that $\rchi(\pm i)=\pi/(2\alpha)$.
 
\paragraph{Baxter equation.} It is convenient to rewrite the relation \re{Q-op} as
\begin{align}\label{pre-TQ}
\left[1- \left(\lambda \bm K+{i\beta\over\pi}\bm H \right)\bm\rchi\right](Q_\pm(x) e_+(x))=e_\pm(x)\,.
\end{align}
This is the operator form of the Fredholm equation \re{Q-defs}. Note that, in contrast to \re{Q-direct-continuation}, the $H-$integral in \re{Q-defs} is understood using the principal value.

Let us apply the operator $4\cosh^2\bm p$ to both sides of \re{pre-TQ}. This operator annihilates $e_\pm(x)$ and its action on $Q_\pm(x) e_+(x)$ is  
\begin{align}
  4\cosh^2\bm p\,(Q_\pm(x) e_+(x))= -\left[Q_\pm(x+2i)+Q_\pm(x-2i)-2Q_\pm(x) \right]e_+(x)\,.
\end{align}
Here $Q_\pm (x+2i)$ and $Q_\pm (x-2i)$ denote the boundary values reached from inside the fundamental strip by the Baxter continuation
\re{Q-Baxter-continuation}.
It follows from the momentum-space representation \re{KH-momentum} of the operators that $4\cosh^2\bm p\,  \bm K=4\pi$ and 
$4\cosh^2\bm p\,(i\bm H)=4 \sinh(2\bm p)$. Applying these relations we find from \re{pre-TQ}
\begin{align}\label{Q-TQ}
\Delta_+(x+2i)Q_\pm (x+2i)
+\Delta_-(x-2i)Q_\pm (x-2i)
=\Big[2-4\pi\lambda\rchi(x)\Big]Q_\pm(x)\,,
\end{align}
where the dressing factors are defined in \re{Q-dressing-factors}. For equal masses, $\beta=0$, the dressing factors reduce to unity. 
The equation \re{Q-TQ} holds for real $x$ and it takes the form  of a  Baxter TQ--equation.  
The  solutions to \re{Q-TQ} are selected by imposing the additional conditions of  analyticity and growth at infinity.
 
\paragraph{Spectral decomposition.} Replacing the inverse spectral operator in \re{Q-op} with its spectral representation \re{dec} gives
\begin{align}\label{Q-sum}
Q_\pm(x)={1\over e_+(x)\rchi(x)}
\sum_{n\ge0}\psi_n(x)
{\vev{\psi_n|e_\pm}\over1-\lambda\mu_n}\,.
\end{align}
It follows from this relation that $Q_\pm(x)$ are meromorphic in $\lambda$, with possible simple poles
at $\lambda=1/\mu_n$. Moreover, the selection rules \re{sec} imply that $Q_+(x)$ receives contributions
only from the even-labelled sector,
\begin{align}
Q_+(x)={1\over e_+(x)\rchi(x)}
\sum_{n\ge0}\psi_{2n}(x)
{\vev{\psi_{2n}|e_+}\over1-\lambda\mu_{2n}}\,.
\end{align}
For $\beta\ne0$, the function $Q_-(x)$ generally receives contributions from both
sectors. 

It proves convenient to introduce the functions
\begin{align}\label{Q-odd-combination}\notag
{}& \mathcal Q_+(x)=Q_+(x)\,,
\\
{}& \mathcal Q_-(x)=Q_-(x)+{i\beta\over\pi^2\lambda}Q_+(x)\,.
\end{align}
The rationale for taking the linear combination in the second relation is that even-sector poles cancel in $\mathcal Q_-(x)$.
Indeed, taking into account the even-sector relation in \re{sec} we find that the even-sector denominators cancel in
\re{Q-odd-combination}, yielding
\begin{align} 
\mathcal Q_-(x)={1\over e_+(x)\rchi(x)}\biggl[
&\sum_{n\ge0}\psi_{2n+1}(x)
{\vev{\psi_{2n+1}|e_-}\over1-\lambda\mu_{2n+1}}
 +{i\beta\over\pi^2\lambda}
\sum_{n\ge0}\psi_{2n}(x)\vev{\psi_{2n}|e_+}\biggr]\,.
\end{align}
The second sum can be extended to the full spectrum, because the
odd-sector overlap with $e_+$ vanishes. Completeness in the inner
product \re{inner}, or equivalently the relation \re{dec} at $\lambda=0$, then gives
\begin{align}\label{Q-odd-sum}
\mathcal Q_-(x)={1\over e_+(x)\rchi(x)}\biggl[
&\sum_{n\ge0}\psi_{2n+1}(x)
{\vev{\psi_{2n+1}|e_-}\over1-\lambda\mu_{2n+1}}
 +{i\beta\over\pi^2\lambda}\bra{x}\bm M^{-1}\ket{e_+}
\biggr]\,.
\end{align}
Apart from the explicit pole at $\lambda=0$, this combination has possible poles only at
the odd-labelled spectral values. The pole at the origin is introduced
by the definition \re{Q-odd-combination} and does not correspond to a
bound-state level. 

Thus, the functions $\mathcal Q_+(x)$ and $\mathcal Q_-(x)$ only have spectral poles in $\lambda$ in the even and odd sectors, respectively.
These sector-dependent pole structures will be
essential when relating the $Q$-functions to $D_+(\lambda)$ and $D_-(\lambda)$.
 
\paragraph{Relation to the potentials.}
We use the definition of the potentials \re{V++} to write
\begin{align}
V_{\pm\pm}(\lambda) = \int_{-\infty}^\infty dx\, e_\pm(x) \vev{x|\bm \rho^{-1}(\lambda,\beta)|e_\pm}\,.
\end{align}
We observe that the same matrix element enters the definition \re{Q-op} of the functions $Q_\pm(x)$. In this way, we obtain
\begin{align}\notag\label{V-Q}
{}& V_{++}(\lambda) = \int_{-\infty}^\infty dx\, e^2_+(x)\rchi(x) Q_+(x)\,,
\\
{}& V_{--}(\lambda) = \int_{-\infty}^\infty dx\, e_+(x)e_-(x)\rchi(x) Q_-(x)\,.
\end{align}
In the next subsection we encounter two additional potentials  
\begin{align}\label{V+-}\notag
{}&V_{+-}(\lambda) = \vev{e_+ |\bm \rho^{-1}(\lambda,\beta)|e_-}=\sum_{n\ge 0} {\vev{e_+|\psi_n}\vev{\psi_n|e_-}\over 1-\lambda\mu_n}\,,
\\
{}&V_{-+}(\lambda) = \vev{e_- |\bm \rho^{-1}(\lambda,\beta)|e_+}=\sum_{n\ge 0} {\vev{e_-|\psi_n}\vev{\psi_n|e_+}\over 1-\lambda\mu_n}\,.
\end{align} 
They are complex conjugates for real parameters and have integral representations analogous to \re{V-Q}
\begin{align}\notag\label{V-Q1}
{}& V_{+-}(\lambda) = \int_{-\infty}^\infty dx\, e^2_+(x)\rchi(x) Q_-(x)\,,
\\
{}& V_{-+}(\lambda) = \int_{-\infty}^\infty dx\, e_+(x)e_-(x)\rchi(x) Q_+(x)\,.
\end{align}
It follows from \re{psi-e} that the potentials \re{V+-} satisfy
\begin{align}\label{V-rel1}
V_{+-}(\lambda)=\widebar V_{-+}(\lambda)=-V_{-+}(\lambda)\,.
\end{align}
Hence $V_{-+}(\lambda)$ is purely imaginary for real $\lambda$
away from the spectral poles. 

The mixed potentials \re{V+-} are not independent. They can be expressed in terms of the potential $V_{++}(\lambda)$. Applying
\re{sec} to their spectral representation \re{V+-} gives
\begin{align}
V_{-+}(\lambda)=\sum_{n\ge 0} {\vev{e_-|\psi_{2n}}\vev{\psi_{2n}|e_+}\over 1-\lambda\mu_{2n}}={i\beta\over\pi^2}\sum_n {\mu_{n}\over 1-\lambda \mu_{n}}|\vev{e_+|\psi_{n}}|^2\ .
\end{align}
 Comparing the last relation with \re{V++} we arrive at
\begin{align}\label{V-rel2}
V_{-+}(\lambda)={i\beta\over\pi^2\lambda}\Big[V_{++}(\lambda)-V_{++}(0)\Big]\,.
\end{align}

\subsection{The first localization identity}

Let us differentiate both sides of \re{Q-direct-continuation} with respect to $z$ and set $z=i$. Since $1/e_+(z)=\cosh(\pi z/2)$ vanishes for $z=i$, the term involving the derivative of the kernel drops out, leaving  
\begin{align}\label{Bax-der}
\widetilde Q^{\,\prime}_\pm(i)- {i\pi \over 2}\int_{-\infty}^\infty dy \left(\lambda  K(i-y)+{i\beta\over\pi}H(i-y) \right)e_+(y)\rchi(y) Q_\pm(y) =E^{\,\prime}_\pm(i)\ ,
\end{align}
where the inhomogeneous terms are $E^{\,\prime}_+(i)=0$ and $E^{\,\prime}_-(i)=1$. Primes on these functions denote derivatives with respect to their
argument at fixed $\lambda$.   

The kernels entering the relation \re{Bax-der} can be expressed in terms of the reference states \re{epm} 
\begin{align}\notag
{}& K(i-y)={\pi  (1+i y)\over 2\cosh\left(\frac{\pi  y}{2}\right)}={\pi\over 2}[e_+(y)+i e_-(y)]\,,
\\
{}&  H(i-y)=-{i\over \cosh({\pi y\over 2})}=-i e_+(y)\,.
\end{align}
Substituting these expressions into \re{Bax-der} we find that the integral on the left-hand side of \re{Bax-der} reduces to the potentials \re{V-Q} and \re{V-Q1} 
\begin{align}\notag\label{wQ}
{}& \widetilde Q^{\,\prime}_+(i)={i\pi\over 2}\left[ {\pi\over 2}\lambda \Big(V_{++}(\lambda)+iV_{-+}(\lambda)\Big)+{\beta\over\pi}V_{++}(\lambda)\right],
\\
{}& \widetilde Q^{\,\prime}_-(i)=1+{i\pi\over 2}\left[ {\pi\over 2}\lambda \Big(V_{+-}(\lambda)+iV_{--}(\lambda)\Big)+{\beta\over\pi}V_{+-}(\lambda)\right].
\end{align}
Applying  \re{V-rel1} and \re{V-rel2}, the right-hand side of both relations can be expressed in terms of the potentials $V_{++}(\lambda)$ and $V_{--}(\lambda)$.

We recall that the Baxter functions $Q_\pm(z)$  are related to the functions $\widetilde Q_\pm(z)$ through the matching relations \re{Q-Baxter-continuation}. We have seen that for real $x$ the analytic properties of the Baxter functions in $\lambda$ simplify after switching from the functions $Q_\pm(x)$ to their linear combinations \re{Q-odd-combination}. Applying 
\re{Q-Baxter-continuation} we can analytically continue $\mathcal Q_\pm(x)$ to complex $x$.
In particular, it follows from \re{Q-i} and \re{Q-odd-combination} that
\begin{align} \label{sp}
\mathcal Q_+(i)
&=\frac{\alpha}{\alpha+\beta}\,,\qqqquad
\mathcal Q_-(i)
=i\,\frac{\alpha}{\alpha+\beta}
\left(1+\frac{\beta}{\pi^2\lambda}\right)\,.
\end{align}
In addition, the derivatives $\widetilde Q^{\,\prime}_\pm(i)$ are related to $\mathcal Q^{\,\prime}_\pm(i)$ by a linear transformation
\begin{align}\notag
\widetilde Q_+'(i)
&={\alpha+\beta\over\alpha}\,\mathcal Q_+'(i)
-{i\beta\pi^2\over4\alpha(\alpha+\beta)}\,,
\\
\widetilde Q_-'(i)
&={\alpha+\beta\over\alpha}
\left[\mathcal Q_-'(i)-{i\beta\over\pi^2\lambda}\mathcal Q_+'(i)\right]
+{\beta\pi^2\over4\alpha(\alpha+\beta)}\,.
\end{align}
Substituting these relations into \re{wQ},  we can solve the resulting system and express the potentials $V_{++}(\lambda)$ and $V_{--}(\lambda)$ in terms of the derivatives $\mathcal Q^{\,\prime}_\pm(i)$
\begin{align}\notag\label{V-localized1}
V_{++}(\lambda)
&=-{4i(\alpha+\beta)\over\alpha(\pi^2\lambda+\beta)}
\mathcal Q_+'(i)
-{\beta\over\pi^2\lambda+\beta}
\left[V_{++}(0)+{\pi^2\over\alpha(\alpha+\beta)}\right]\,,
\\\notag
V_{--}(\lambda)
&={4\over\pi^2\lambda}
-{4(\alpha+\beta)\over\alpha\pi^2\lambda}\mathcal Q_-'(i)
-{4i\beta^2(\alpha+\beta)\over
\alpha\pi^4\lambda^2(\pi^2\lambda+\beta)}\mathcal Q_+'(i)
\\ 
&\quad
-{\beta\bigl(\pi^4\lambda^2+2\beta\pi^2\lambda+2\beta^2\bigr)
\over\alpha(\alpha+\beta)\pi^2\lambda^2(\pi^2\lambda+\beta)}
-{\beta(\pi^2\lambda+2\beta)^2\over
\pi^4\lambda^2(\pi^2\lambda+\beta)}V_{++}(0)\,.
\end{align}
Thus the potentials $V_{++}(\lambda)$ and $V_{--}(\lambda)$ are expressed through derivatives
of the Baxter functions $\mathcal{Q}_\pm(z)$ at $z=i$ and the reference value
$V_{++}(0)$. Notice that  $\mathcal Q^{\,\prime}_\pm(i)$ depend on the spectral parameter $\lambda$ and the mass parameters $\alpha$ and $\beta$.

Substituting \re{V-localized1} into \re{dD-} and \re{dD+}  gives the localization identities for the spectral determinants in the two sectors. Using the special-point values of $\mathcal Q_\pm$ in \re{sp}, we can write the result in terms of logarithmic derivatives of $\mathcal Q_\pm(x)$ at $x=i$
\begin{align}\notag\label{localization-Dminus}
\partial_\lambda\log D_-(\lambda)
{}&= \left(1+{\beta\over\pi^2\lambda}\right)\bigg[2i
(\log\mathcal Q_-(i))^\prime +{\beta\pi^2\over2\alpha(\alpha+\beta)}\bigg]
\\ 
{}&+\left({\beta\over2}+{\beta^2\over\pi^2\lambda}\right)V_{++}(0)
-{\beta^2\over2\pi^2}V_{++}'(0)+d_--2
 \,,
\\[2mm]\notag\label{localization-Dplus}
\partial_\lambda\log D_+(\lambda)
{}&={\bigl(\pi^4\lambda^2-2\alpha\pi^2\lambda+\beta^2\bigr)
\over\pi^2\lambda(\pi^2\lambda+\beta)}\bigg[
2i(\log\mathcal Q_+(i))^\prime+{\beta\pi^2\over2\alpha(\alpha+\beta)}\bigg]
\\ 
&+\left[{\beta\over2}
-{\alpha(\pi^2\lambda+2\beta)\over\pi^2\lambda+\beta}
+{\beta^3\over\pi^2\lambda(\pi^2\lambda+\beta)}\right]V_{++}(0)
+{\beta^2\over2\pi^2}V_{++}'(0)+d_+ \,.
\end{align}
Here $V_{++}'(0)$ denotes differentiation with respect to $\lambda$ and the notation was introduced for the logarithmic derivatives of $\mathcal Q_\pm(x)$ with
respect to $x$ at fixed $\lambda$ 
\begin{align}
(\log\mathcal Q_\pm(i))^\prime=\partial_x\log\mathcal Q_\pm(x)\Big|_{x=i}.
\end{align}
The relations \re{localization-Dminus} and \re{localization-Dplus} hold for arbitrary values of the mass parameters $\alpha$ and $\beta$.  For $\beta=0$, they reduce to
\begin{align}\notag\label{localization-equal-masses}
\partial_\lambda\log D_-(\lambda)
&=2i\,(\log\mathcal Q_-(i))^\prime-2+d_-\,,
\\[2mm]
\partial_\lambda\log D_+(\lambda)
&=2i\left(1-{2\alpha\over\pi^2\lambda}\right)
(\log\mathcal Q_+(i))^\prime
-\alpha V_{++}(0)+d_+\,.
\end{align}
The  identities \re{localization-Dminus} and \re{localization-Dplus} involve four $\lambda$-independent parameters: the subtraction constants $d_\pm$ and the first two coefficients in the small-$\lambda$ expansion of the potential, $V_{++}(0)$ and $V_{++}'(0)$. These parameters are determined in Appendix~\ref{app}. 

As noted above, the constants $d_\pm$ depend on the subtraction convention and may therefore be chosen for convenience. In the next subsection, we relate the ratio of the regularized spectral determinants to the ratio of the Baxter $Q-$functions evaluated at special points. We fix the subtraction scheme by requiring this relation to take a simple form.

Repeating the calculation of \re{Bax-der} at $z=-i$ expresses the logarithmic
derivatives of $D_\pm(\lambda)$ through derivatives of the Baxter
functions at $x=-i$. The relation between the two 
representations follows by combining \re{wQ} with its counterpart
at $z=-i$ and applying the matching conditions
\re{Q-Baxter-continuation}
\begin{align}\label{localization-special-point-relation}
\left(1\pm\frac{\beta}{\pi^2\lambda}\right)(\log\mathcal Q_\pm(-i))'{}&+\left(1\mp\frac{\beta}{\pi^2\lambda}\right)(\log\mathcal Q_\pm(i))'
=\frac{i\beta}{2}\left[V_{++}(0)+\frac{\pi^2}{\alpha^2-\beta^2}\left(1\pm\frac{\beta^2}{\pi^2\alpha\lambda}\right)\right],
\end{align}
where the upper and lower signs refer to $\mathcal Q_+$ and $\mathcal Q_-$, respectively. Using \re{localization-special-point-relation} to
eliminate the derivatives $(\log\mathcal Q_\pm(i))'$ in \re{localization-Dminus} and \re{localization-Dplus} gives the equivalent identities involving $(\log\mathcal Q_\pm(-i))'$. At equal masses, $\beta=0$, the parity relations $\mathcal Q_\pm(-x)=\pm\mathcal Q_\pm(x)$ imply that both logarithmic derivatives are odd and, therefore, equation \re{localization-special-point-relation} is satisfied identically.
 
\subsection{The second localization identity}

The first localization identity, \re{localization-Dminus} and \re{localization-Dplus}, expresses the logarithmic derivatives
of $D_\pm(\lambda)$ through derivatives of the Baxter functions at
$x=i$. In this subsection, we derive the second identity which relates the ratio of the two determinants to the values
of $\mathcal Q_\pm(x|\lambda)$ at $x=0$. For clarity, we display the dependence of the Baxter $Q-$functions on $\lambda$.

The spectral representations derived above show that
$\mathcal Q_+(x|\lambda)$ has poles in $\lambda$ only at the even-labelled
levels $\lambda_{2n}$, whereas $\mathcal Q_-(x|\lambda)$ has 
spectral poles in $\lambda$ only at the odd-labelled levels $\lambda_{2n+1}$.
The latter also contains the explicit pole at $\lambda=0$ introduced
by \re{Q-odd-combination}. Thus the products
$\mathcal Q_+(x|\lambda)D_+(\lambda)$ and
$\lambda\mathcal Q_-(x|\lambda)D_-(\lambda)$ are entire in $\lambda$. We show in this subsection that for $x=0$ these products are proportional to $D_-(\lambda)$ and $D_+(\lambda)$, respectively. 

\paragraph{Determinant representations.} 
Evaluating \re{Q-op} and \re{Q-odd-combination}
at $x=0$ gives
\begin{align}\label{Q_+(0)} 
\mathcal Q_+(0|\lambda){}&=\bra{0}\Big[1-\Big(\lambda \bm K+{i\beta\over\pi} \bm H\Big)\bm\rchi\Big]^{-1}\ket{e_+}
=1+\bra{e_0}\bm\rho^{-1}\ket{e_+}
=\det\big( \bm\rho^{-1}\bm\rho_+\big)\,,
\end{align}
where we added and subtracted $1$ from the operator inside the matrix element in the first relation and used $e_+(0)=1$ and $e_-(0)=0$.
Here the spectral operator $\bm \rho$ is given by \re{rho} and the notation was introduced for  
\begin{align}\label{rho+} 
\bm\rho_+=\bm\rho + \ket{e_+}\bra{e_0}\,,\qqqquad \bra{e_0}=\bra{0}\Big(\lambda \bm K+{i\beta\over\pi} \bm H\Big)\ .
\end{align}
The same calculation of $\mathcal Q_-(0)$ gives 
\begin{align}\label{Q_-(0)}
\mathcal Q_-(0|\lambda)={i\beta\over \pi^2\lambda}\left[1+\bra{e_0}\bm\rho^{-1}\ket{e_++{ \pi^2\lambda\over i\beta}e_-}\right]
={i\beta\over \pi^2\lambda}\det\big(\bm\rho^{-1}\bm\rho_-\big) \, ,
\end{align}
where the operator $\bm\rho_-$ is defined as
\begin{align}\label{rho-} 
\bm\rho_-=\bm\rho+\lr{\ket{e_+}+{\pi^2\lambda\over i\beta} \ket{e_-}}\bra{e_0}\,.
\end{align}
The determinants in \re{Q_+(0)} and \re{Q_-(0)} are relative
determinants of rank-one perturbations. After the same spectral
regularization as in Section~\ref{sect2}, their common denominator has the
zeros of $D(\lambda)=D_+(\lambda)D_-(\lambda)$.
The absence of odd-sector poles in $\mathcal Q_+(0|\lambda)$ requires
the numerator associated with $\bm\rho_+$ to vanish at the zeros of
$D_-$. Similarly, the numerator associated with $\bm\rho_-$ must
vanish at the zeros of $D_+$. We show below that these numerator
zeros are double. 
  
The operators \re{rho+} and \re{rho-} differ from $\bm\rho$ by a rank-one perturbation.  These operators are non-Hermitian,
so their left and right null vectors have to be treated separately. It follows from the commutation relations \re{H-com} and the definition \re{rho+} of the state $\bra{e_0}$ that the operators $\bm\rho$ and $\bm\rho_\pm$ satisfy the relations
\begin{align}\notag\label{mod-com}
{}&[\bm\rho,\bm t]={\pi\lambda\over 2}\lr{\ket{e_-}\bra{e_+}-\ket{e_+}
\bra{e_-}}+{i\beta\over\pi}\ket{e_+}\bra{e_+}\,,
\\
\notag
{}&\bm\rho_+\bm t=\bm t\bm\rho+\frac{\pi\lambda}{2}\ket{e_-}\bra{e_+}\,,
\\
{}&\bm\rho_- \bm t=\bm t \bm\rho+{\pi\lambda\over 2}\ket{e_-}\lr{{\pi^2\lambda\over i\beta}\bra{e_-}-\bra{e_+}}\,.
\end{align}
Here the multiplication operator $\bm t=\tanh(\pi\bm x/2)$ is introduced in \re{Q=t}. Notice that the bra vectors in the last two relations are precisely those selected
by the two sector conditions \re{sec} (for $\lambda=1/\mu_{2n}$). This explains the choice of the rank-one perturbation in \re{rho+} and \re{rho-}. 

For the odd-labelled states, we use \re{eig} combined with \re{dec} and \re{mod-com} to verify that 
\begin{align}\notag\label{eig+}
{}&\bm\rho_+\bm t\ket{\psi_{2n+1}}= (1-\lambda\mu_{2n+1})\bm M\bm t\ket{\psi_{2n+1}} \,,
\\[2mm]
{}&\bra{\psi_{2n+1}} \bm\rho_+=(1-\lambda \mu_{2n+1}) \bra{\psi_{2n+1}}\bm M\,,
\end{align} 
where the operator $\bm M$ is defined in \re{M}. At $\lambda=\lambda_{2n+1}=1/\mu_{2n+1}$, the corresponding right and left
null vectors are $\bm t\ket{\psi_{2n+1}}$ and $\bra{\psi_{2n+1}}$. Most importantly, these null vectors are orthogonal with respect to the $\bm M$-weighted inner
product \re{inner}
\begin{align}\label{off-diag}
\vev{\psi_{2n+1}|\bm M \bm t|\psi_{2m+1}} = 0
\end{align}
for arbitrary $m$ and $n$.
For $m\neq n$ this relation follows from \re{eig+} by comparing the two expressions for $\bra{\psi_{2n+1}}\bm\rho_+\bm t\ket{\psi_{2m+1}}$. For $m=n$ the matrix element in \re{off-diag} vanishes because the $\bm{CP}$ symmetry makes it  purely imaginary,
whereas $[\bm M,\bm t]\ket{\psi_{2n+1}}=0$ makes it real.
 
Choosing an $\bm M$-orthogonal basis $\ket{u_1}=\bm t\ket{\psi_{2n+1}}$ and $\ket{u_2}=\ket{\psi_{2n+1}}$, we can define a matrix $\vev{u_i|\bm\rho_+|u_j}$. 
This matrix is triangular: its diagonal matrix elements are proportional to $(1-\lambda\mu_{2n+1})$, while the off-diagonal matrix element $\vev{u_2|\bm\rho_+|u_1}$ vanishes by virtue of \re{off-diag}. As a consequence, the determinant of this matrix is proportional to $(1-\lambda\mu_{2n+1})^2$, so that the regularized determinant associated with $\bm\rho_+$ has a double zero at $\lambda=1/\mu_{2n+1}$ and, therefore, is proportional to $D_-^2(\lambda)$.
 
The same calculation shows that the operator $\bm\rho_-$ satisfies the relation analogous to \re{eig+}
\begin{align}\notag
{}&\bm\rho_- \bm t\ket{\psi_{2n}}=(1-\lambda\mu_{2n})\Big[\bm M \bm t \ket{\psi_{2n}}-\bm t\ket{e_0}\vev{e_+|\psi_{2n}}\Big]\,,
\\ 
{}&\bra{\psi_{2n}}\bm\rho_-=(1-\lambda\mu_{2n}) \Big[\bra{\psi_{2n}}\bm M+\vev{\psi_{2n}|e_+} \bra{e_0}\Big]\,,
\end{align} 
where the eigenstates $\psi_{2n}(x)$ in the even sector satisfy \re{sec}. 
In contrast to \re{off-diag}, the corresponding $\bm M-$weighted matrix elements do not vanish
\begin{equation}
\langle\psi_{2n}|\bm M\bm t|\psi_{2m}\rangle
=\frac{i\beta}{2\pi}
\langle\psi_{2n}|e_+\rangle
\langle e_+|\psi_{2m}\rangle \,.
\end{equation}
However, the matrix elements $\bra{\psi_{2n}}\bm\rho_- \bm t\ket{\psi_{2m}}$ are given by 
\begin{align}
\bra{\psi_{2n}}\bm\rho_- \bm t\ket{\psi_{2m}} =-{i\beta\over 2\pi} (1-\lambda\mu_{2n})(1-\lambda\mu_{2m})\vev{\psi_{2n}|e_+}\vev{e_+|\psi_{2m}}\,.
\end{align}
This ensures that the regularized determinant associated with $\bm\rho_-$ has a double zero at $\lambda=1/\mu_{2n}$ and, therefore, is proportional to $D_+^2(\lambda)$.

\paragraph{Normalization and subtraction scheme.} 
In the determinant representations \re{Q_+(0)} and \re{Q_-(0)},
the regularized numerators contain the factors $D_-^2$ and
$D_+^2$, respectively, and the common denominator has the product
$D_+D_-$. Cancelling the common factors leaves inverse ratios of the
sector determinants. With the subtraction convention of
Section~\ref{sect2}, the remaining factors are a constant and an
exponential linear in $\lambda$ for each ratio,
\begin{align}\notag\label{Q-spec}
{}&\mathcal Q_+(0|\lambda)=\mathcal N_+\, e^{\lambda \delta_+}{D_-(\lambda)\over D_+(\lambda)}\,,
\\
{}&\mathcal Q_-(0|\lambda)= \mathcal N_-\, e^{\lambda \delta_-} {i\beta\over \pi^2\lambda}{D_+(\lambda)\over D_-(\lambda)}\,.
\end{align}
Here $\mathcal N_\pm$ are normalization factors, and the constants
$\delta_\pm$ compensate the subtraction convention for $D_\pm(\lambda)$.

Taking the value and logarithmic derivative of the first relation
in \re{Q-spec} at $\lambda=0$, with $D_\pm(0)=1$ and
$\partial_\lambda\log D_\pm(0)=d_\pm$, gives
\begin{align} 
  \mathcal N_+=\mathcal Q_+(0|0)\,, 
 \qqquad \delta_+=\Delta-d_-+d_+\,,
\qqquad \Delta= \partial_\lambda
 \log\mathcal Q_+(0|\lambda)\Big|_{\lambda=0}\,.
\end{align}
The remaining parameters  $\delta_-$ and $\mathcal N_-$ follow from the relation 
\begin{align}\label{reciprocal}\notag
{}& \mathcal Q_+(0|\lambda)\mathcal Q_-(0|\lambda)
 =\frac{i\beta}{\pi^2\lambda}
 [\mathcal Q_+(0|0)]^2\,,
 \\
 {}&\mathcal Q_+(0|0)=\bigg[1-{4\beta^2\rchi^2(0)\over \pi^2}\bigg]^{-1/2}={\alpha+1\over\gamma}\, ,
\end{align}
where  $\gamma=\sqrt{(1+\alpha)^2-\beta^2}$. The derivation of (\ref{reciprocal}) can be found in Appendix~\ref{app}. Together with \re{Q-spec} it leads to $\mathcal N_+=\mathcal N_-$ and $\delta_-=-\delta_+$.
Combining these results we arrive at the second localization identity
\begin{align}\notag\label{second}
{}&\mathcal Q_+(0|\lambda)=e^{\lambda (\Delta-d_-+d_+)} \mathcal Q_+(0|0) {D_-(\lambda)\over D_+(\lambda)}\,,
\\
{}&\mathcal Q_-(0|\lambda)= e^{-\lambda (\Delta-d_-+d_+)} \mathcal Q_+(0|0) {i\beta\over \pi^2\lambda}{D_+(\lambda)\over D_-(\lambda)}\,.
\end{align}
The subtraction constants $d_\pm$ encode the freedom in defining the regularized determinants $D_\pm(\lambda)$. 
The Baxter functions on the left-hand side of \re{second} are independent of this convention. We can use this property to choose 
\begin{align}\label{d-diff}
d_--d_+=\Delta\,.
\end{align}
In this scheme, the relation \re{second} takes the form
\begin{align}\notag\label{second1}
{}&\mathcal Q_+(0|\lambda)=\frac{\alpha+1}{\gamma} {D_-(\lambda)\over D_+(\lambda)}\,,
\\
{}&\mathcal Q_-(0|\lambda)= \frac{\alpha+1}{\gamma} {i\beta\over \pi^2\lambda}{D_+(\lambda)\over D_-(\lambda)}\,,
\end{align}
where we replaced the symbol function with its expression \re{chi-QCD}. 

At equal masses, for $\beta=0$, the first identity in \re{second1} becomes
$\mathcal Q_+(0|\lambda)=D_-(\lambda)/D_+(\lambda)$, while the second gives
$\mathcal Q_-(0|\lambda)=0$, consistent with the parity relation $\mathcal Q_-(-x|\lambda)=-\mathcal Q_-(x|\lambda)$.

\paragraph{Comparison with the conjectured identities.}
The relations \re{localization-Dminus}, \re{localization-Dplus} and \re{second1}
can be compared with the conjectures of
Ref.~\cite{Artemev:2025cev}, provided the definition of the Baxter $Q$-functions and the
determinant normalizations are matched.
 
The functions denoted by $\bf Q_+$ and $\bf Q_-$ in
that reference correspond to $\mathcal Q_-$ and $\mathcal Q_+$,
respectively, up to factors independent of $x$
\begin{align}\label{Q-bf}
{\bf Q}_-(x) = {\mathcal Q_+(x|\lambda)\over \mathcal Q_+(0|\lambda)}\,,\qqqquad 
\mathbf Q_+(x)
=\frac{2i\beta(1+\alpha)}
{\pi^2\lambda(1+\alpha+\gamma)}
\frac{\mathcal Q_-(x|\lambda)}{\mathcal Q_-(0|\lambda)}\,.
\end{align}
The additional factors do not affect the logarithmic $x-$derivatives entering the localization identities. 

At equal masses, for $\beta=0$, the relation \re{localization-equal-masses} reproduces
Eqs.~(5.16) and (5.17) of Ref.~\cite{Artemev:2025cev}. For $\beta\neq 0$, 
the relations \re{localization-Dminus} and \re{localization-Dplus} correspond, respectively, to Eqs.~(5.18) and (5.21) (or equivalently
Eq.~(5.22))
of Ref.~\cite{Artemev:2025cev}.
   
Finally, setting $x=i$ in \re{Q-bf} and using the relations \re{sp} and \re{second1} give
\begin{align}\notag 
{}& {\bf Q}_-(i) = \frac{\alpha}{\alpha+\beta}{\gamma\over 1+\alpha} {D_+(\lambda)\over D_-(\lambda)}\,,
\\
{}& {\bf Q}_+(i) =2i \frac{\beta+\pi^2\lambda}{\pi^2\lambda} \frac{\alpha}{\alpha+\beta}\frac{\gamma}{1+\alpha+\gamma}{D_-(\lambda)\over D_+(\lambda)}\,.
\end{align}
These are the determinant-ratio relations conjectured in
Eq.~(5.15) of Ref.~\cite{Artemev:2025cev}. 

\section{Localization for general symbols}
\label{sec:rational-symbols}

In the previous section, we derived the localization identities
\re{localization-Dminus} and \re{localization-Dplus} for the sector
determinants $ D_\pm(\lambda)$ with the QCD symbol \re{chi-QCD}.
These identities express $\partial_\lambda\log D_\pm(\lambda)$ through the derivatives of the Baxter $Q-$functions at $x=i$. 
The integral representation \re{intro:dlogD}, which arises in the
Tracy--Widom approach, suggests that the localization mechanism might follow from
contour deformation. However, both the QCD symbol and the Baxter
$Q$-functions are analytic at $x=i$ for generic parameter values.
Consequently, this point does not provide a pole and the origin of the mechanism must instead be understood within the spectral
problem.
This observation leads to three questions:
\begin{itemize}
\item Which properties of the QCD symbol \re{chi-QCD} select $x=i$
as the evaluation point in \re{localization-Dminus} and
\re{localization-Dplus}?
\item For which symbols $\rchi(x)$ do analogous localization
identities hold?
\item Which properties of the integral operators $\bm K$ and $\bm H$
are required for this construction?
\end{itemize}
In this section, we address the first two questions. We keep the operators $\bm K$ and $\bm H$ fixed and we generalize the symbol $\rchi$. 

\subsection{Spectral sectors and rational closure}

In order to generalise the structure of $\rchi(x)$, it is convenient to start from  the QCD symbol \re{chi-QCD} and express it by the $t$-variable 
\begin{align}\label{x(t)}
t=\tanh\left(\frac{\pi x}{2}\right)\, , \qqqquad x(t)=\frac1\pi\log\frac{1+t}{1-t}\, .
\end{align}
The real $x-$axis maps to the open interval $-1<t<1$, with the endpoints corresponding to $x\to\pm\infty$. By the $2i$ periodicity of $t(x)$, the boundary lines $\Im x=\pm2$ of the fundamental strip map to the same interval. In the $t$-variable, the QCD symbol \re{chi-QCD} obeys
\begin{align}\label{QCD-prod}
t \,\rchi^{-1}(x) = x + {2\alpha\over\pi}t\, .
\end{align}

The two virial identities \re{vir-eq1} and \re{qcd} play complementary roles in the derivation of the localization identities. The first one follows from the commutation relations \re{H-com} and the fact that $\bm t$ commutes with the symbol. Together with the $\bm C \bm P$-symmetry, it yields the selection rules \re{sec}, which remain valid for more general symbols provided the same symmetry and spectral assumptions hold. 
Conversely, the second identity \re{aux2} requires the explicit form of the function $\rchi(x)$.

These observations suggest considering the following general real even symbol function $\rchi(x)$ 
\begin{align}\label{chi-gen}
A(t) \,\rchi^{-1}(x)=x+B(t)\, ,
\end{align}
where $t$ is given by \re{x(t)}, $A(t)$ and $B(t)$ are real odd rational functions with at most linear
growth at infinity and finitely many simple poles
\begin{align}
	\label{AB}
 A(t)&=c_\infty t+\sum_{a=1}^N\frac{c_a}{t_a-t}\, ,\notag\\
  B(t)&=d_\infty t+\sum_{a=1}^N\frac{d_a}{t_a-t}\, . 
\end{align}In the previous expression,  $c_\infty$, $d_\infty$ are real parameters, while oddness and reality imply 
\begin{equation}
	c_{-a}=c_a\, , \qqquad c_{\bar a}=\overline{c_a}\, , \qquad	d_{-a}=d_a\, , \qqquad d_{\bar a}=\overline{d_a}\, .
\end{equation} For instance, the QCD symbol \re{QCD-prod} corresponds to $c_\infty=1$, $d_\infty=2\alpha/\pi$
and $d_a=c_a=0$. The poles $t_a\notin[-1,1]$ may be complex and belong to a
quartet $(t_a,-t_a,\bar t_a,-\bar t_a)$, which reduces to a pair $(t_a,-t_a)$ on the real or imaginary axis. The sums in \re{AB} include
each distinct pole once.~\footnote{The functions $A(t)$ and $B(t)$ need not share the same poles. By choosing some of the parameters $c_a$ and $d_a$ equal to zero, we can modify the pole structure of the functions.}  These conditions make $A(t)$
and $B(t)$ real and regular for $t\in[-1,1]$, while positivity and spectral assumptions on $\rchi$ are imposed separately. Finally, we observe that higher-order poles can be obtained by coalescing
simple poles with suitably scaled coefficients and contribute to the localization formulas by higher derivatives of the Baxter $Q$-functions.

\subsection*{Finite-rank commutators and virial generator}

For each pole $t_a$, we define the right and left sources
\begin{equation}
	\label{fivebis:sources}
	|e_a^R\rangle=(t_a-\bm t)^{-1}|e_+\rangle\, ,
	\qquad
	\langle e_a^L|=\langle e_+|(t_a-\bm t)^{-1}\, .
\end{equation}
These states are not Hermitian adjoints when $t_a$ is complex. A straightforward calculation shows that the 
commutator \re{H-com} is promoted to \begin{equation}\label{fivebis:pole-commutator}
	[(t_a-\bm t)^{-1},\bm H]=|e_a^R\rangle\langle e_a^L|\, .
\end{equation} As we will shortly see, the relation (\ref{fivebis:pole-commutator}) is the operator origin of the localization mechanism for the more general symbol (\ref{chi-gen}).

To  generalize the virial generator \re{q-def} we choose  
\begin{equation}\label{fivebis:generator}
	\bm q=iA(\bm t)\bm H\, .
\end{equation}
Proceeding as in \re{aux2}, we find that the terms proportional to $\beta$ cancel in the combination $\bm\rho\bm q-\bm q^\dagger\bm\rho$, while \re{chi-gen} produces $[\bm x+B(\bm t),\bm H]$. The part involving $\bm x$ gives $(2/\pi)\bm K$, while the rational terms reduce to finite-rank operators \re{fivebis:pole-commutator}.

Taking the expectation value \re{vir} and using the normalization \re{inner}, we obtain
\begin{align}\label{rational-norm-identity}
\begin{aligned}
\mu_n^2={}&
\left(\frac{\pi^2c_\infty}{2}
-\frac{\pi d_\infty\mu_n}{2}\right)
|\langle e_+|\psi_n\rangle|^2
+\frac{\pi^2c_\infty}{2}
|\langle e_-|\psi_n\rangle|^2\\
&+\sum_a \left(\frac{\pi^2c_a}{2}-\frac{\pi d_a\mu_n}{2}\right)\vev{\psi_n|e_a^R}\vev{e_a^L|\psi_n}-\frac{\pi^2}{8}\sum_a c_a\vev{\psi_n|\bm H|e_a^R}\vev{e_a^L|\bm H|\psi_n}\, .
\end{aligned}
\end{align}  Notice that (\ref{rational-norm-identity}) reduces to the QCD relation \re{qcd} in the absence of poles, $c_a=d_a=0$, and for $c_\infty=1$ and $d_\infty=\frac{2\alpha}{\pi}$. Moreover, it is straightforward to verify that $\mu_{n}^2$ is real. \footnote{Indeed, for real $t_a$, the matrix elements in (\ref{rational-norm-identity}) satisfy $ \vev{e_a^L|\psi_n}=\widebar{\vev{\psi_n|e_a^R}}$ and $\vev{e_a^L|\bm H|\psi_n}=-\widebar{\vev{\psi_n|\bm H|e_a^R}}$. For complex $t_a$, these matrix elements are independent. However, the two terms on the first line of \re{rational-norm-identity} are explicitly real, while on the second line the reality is restored in the sum over all poles. }

\subsection{Localization identity}\label{sec:5loc}

Inserting the norm identity \re{rational-norm-identity} into the spectral sum for the full determinant $D(\lambda)$ and applying the resolvent
expansion \re{dec}, we find that  
\begin{equation}
\label{D-W0}
\begin{split}
    \partial_\lambda\log D(\lambda)
&=d-\lambda\sum_{n\ge0}\frac{\mu_n^2}{1-\lambda\mu_n}\\
&=d+f_\infty(\lambda)-f_\infty(0)
+\sum_{a=1}^N\bigl[f_a(\lambda)-f_a(0)\bigr]\, ,
\end{split}
\end{equation}
where $d=d_++d_-$ and the subtractions at $\lambda=0$ implement the regularization, ensuring $\partial_\lambda\log D(0)=d$.

The linear terms in \re{D-W0} are given by a combination of the potentials \re{V++}
\begin{align}\label{W-infty}
f_\infty(\lambda)=\lr{\frac{\pi d_\infty}{2}-\frac{\pi^2\lambda c_\infty}{2}}V_{++}(\lambda)-\frac{\pi^2\lambda c_\infty}{2}V_{--}(\lambda)\, .
\end{align}
Following the approach described in Section~\ref{sec:BaxQCD}, we express the potentials in terms of the Baxter functions $\mathcal{Q}_\sigma(z|\lambda)$ \re{Q-odd-combination}, constructed by \re{Q-defs} with the general symbol \re{chi-gen}. It proves convenient to introduce the dressed functions
\begin{equation}\label{math-Q}
\widetilde{\mathcal Q}_\sigma(z|\lambda)=
\begin{cases}
\Delta_+(z)\mathcal Q_\sigma(z|\lambda),&0<\Im z<2\, ,\\[2mm]
\Delta_-(z)\mathcal Q_\sigma(z|\lambda),&-2<\Im z<0\, ,
\end{cases}
\qquad \sigma=\pm\, ,
\end{equation}
where $\Delta_\pm(z)=1\pm(2\beta/\pi)\rchi(z)$.
 The potentials $V_{++}(\lambda)$ and $V_{--}(\lambda)$ are then
\begin{align}\label{rational-symmetric-potentials}
V_{++}(\lambda)
&=-\frac{2i}{\pi^2\lambda}
\left[\widetilde{\mathcal Q}_+'(i|\lambda)
-\widetilde{\mathcal Q}_+'(-i|\lambda)\right]\, ,
\notag\\[2mm]
V_{--}(\lambda)
&=\frac4{\pi^2\lambda}
-\frac2{\pi^2\lambda}\sum_{\sigma =\pm1}
\left(1-\frac{2\sigma\beta}{\pi^2\lambda}\right)
\left[\widetilde{\mathcal Q}_-'(i\sigma|\lambda)-\frac{i\beta}{\pi^2\lambda}\widetilde{\mathcal Q}_+'(i\sigma |\lambda)\right]\, ,
\end{align}
where a prime denotes differentiation with respect to the first argument at fixed $\lambda$.
Substituting these relations into \re{W-infty}, we  express $f_\infty(\lambda)$ in terms of derivatives of the Baxter functions. 

The contribution of a finite pole $t_a$ to \re{D-W0} enters through 
\begin{align}\label{fa}
f_a(\lambda)=\left(\frac{\pi d_a}{2}-\frac{\pi^2\lambda c_a}{2}\right)\langle e_a^L|\bm\rho^{-1}|e_a^R\rangle
+\frac{\pi^2\lambda}{8} c_a\langle e_a^L|\bm H\bm\rho^{-1}\bm H |e_a^R\rangle\, .
\end{align} 
These matrix elements are evaluated  in Appendix~\ref{app:C}. They involve the Baxter functions $\mathcal{Q}_\sigma(z|\lambda)$ evaluated at two special points in the fundamental strip $-2 < \Im z< 2$. The first one is $x_a$, defined by 
\begin{equation}\label{finite-pole-points}
 t(x_a)=t_a\, ,\qqqquad 0<\Im x_a<2\, ,
\end{equation} while the second one is $x_a-2i$.
Before displaying the explicit expression for the matrix elements, it is useful to introduce notation for 
the sum and difference of the functions \re{math-Q} at these two points 
\begin{align}\notag\label{Q-eps}
{}&\widetilde{\mathcal Q}_\sigma^+(x_a|\lambda)=\widetilde{\mathcal Q}_\sigma(x_a|\lambda)+\widetilde{\mathcal Q}_\sigma(x_a-2i|\lambda)\,,
\\[2mm]   
{}&\widetilde{\mathcal Q}_\sigma^-(x_a|\lambda)=\widetilde{\mathcal Q}_\sigma(x_a|\lambda)-\widetilde{\mathcal Q}_\sigma(x_a-2i|\lambda)\,.
\end{align}
Here the subscript $\sigma=(+,-)$ labels
the Baxter solution, while the superscript `$+$' or `$-$' selects the sum
or difference of its values at $x_a$ and $x_a-2i$.

The first matrix element in \re{fa}, computed in Appendix~\ref{app:C},
takes the form
\begin{equation}\label{1st}
 \langle e_a^L|\bm\rho^{-1}|e_a^R\rangle
 =-\frac{ \Wr(\widetilde{\mathcal Q}_+^-,
                         \widetilde{\mathcal Q}_-^-) }
          {\pi^2\lambda(1-t_a^2)\widetilde{\mathcal Q}_+(x_a|0)}\,,
\end{equation}
where the notation was introduced for the differential Wronskian 
\begin{equation}\label{Wron}
\Wr(\widetilde{\mathcal Q}_\sigma^\epsilon,\widetilde{\mathcal Q}_{\sigma'}^{\epsilon'})={\widetilde{\mathcal Q}}_\sigma^\epsilon(x_a)\partial_{x_a}\widetilde{\mathcal Q}_{\sigma'}^{\epsilon'}(x_a)-\widetilde{\mathcal Q}_{\sigma'}^{\epsilon'}(x_a)\partial_{x_a}\widetilde{\mathcal Q}_\sigma^\epsilon(x_a)\, ,
\end{equation} 
with $\sigma,\sigma'=(+,-)$ and $\epsilon,\epsilon'=(+,-)$.
Here the Baxter functions depend on
the same spectral parameter $\lambda$ and the derivatives act on the common argument $x_a$, at fixed $\lambda$.

Combining \re{1st} with the second matrix element, given in
\re{complex-second-contraction}, yields
\begin{align}\label{eq:finite-pole}
 f_a(\lambda)={}&\frac1{2(1-t_a^2)\widetilde{\mathcal Q}_+(x_a|0)}\bigg\{
 \left(c_a-\frac{d_a}{\pi\lambda}\right)
       \Wr(\widetilde{\mathcal Q}_+^-,
           \widetilde{\mathcal Q}_-^-)
 -c_a\Wr(\widetilde{\mathcal Q}_+^+,
          \widetilde{\mathcal Q}_-^+)
 \notag\\[1mm]
 &+\frac{\beta c_a}{\pi^2\lambda}
       \Big[\Wr(\widetilde{\mathcal Q}_+^+,
                \widetilde{\mathcal Q}_-^-)
          +\Wr(\widetilde{\mathcal Q}_+^-,
               \widetilde{\mathcal Q}_-^+)
          +4i\widetilde{\mathcal Q}_+'(x_a|0)\Big]
 +4c_a\widetilde{\mathcal Q}_+(x_a|0)\bigg\}\, .
\end{align}
The differential Wronskians in this relation depend on $x_a$, while  $\widetilde{\mathcal Q}_+'(x_a|0)$ denotes the derivative of the Baxter function with respect to $x_a$ evaluated at $\lambda=0$. It follows from (\ref{eq:finite-pole}) that the contribution of the pole $t_a$ to \re{D-W0} depends on four Baxter values \re{Q-eps} and their
derivatives. 

Substituting \re{W-infty} and \re{eq:finite-pole} into \re{D-W0}, we can express the logarithmic derivative $\partial_\lambda\log D$ through finitely many values of the Baxter $\mathcal Q$-functions  and their derivatives. The linear terms in \re{AB} select $x=\pm i$, while each finite pole $t_a$ selects the pair of points $x_a$ and $x_a-2i$. 
 
\subsection{Applications}
\label{sec:applications}

We illustrate the localization identity  \re{D-W0} with four examples of the symbol \re{chi-gen}, restricting to $\beta=0$.
The dressing factors $\Delta_\pm(x)$ then reduce to unity. The relations \re{Q-odd-combination} and \re{math-Q} give $\widetilde{\mathcal Q}_\sigma(x)=\mathcal Q_\sigma(x)=Q_\sigma(x)$. The functions $Q_\sigma(x)$ satisfy the inhomogeneous Fredholm equation \re{Q-defs} and, for an even symbol, have definite parity: $Q_\pm(-x|\lambda)=\pm Q_\pm(x|\lambda)$. The spectral determinant factorizes accordingly as $D(\lambda)=D_+(\lambda)D_-(\lambda)$.  

To separate the sectors, we use the second localization identity \re{second}, 
which is independent of the choice of even symbol.
Following \re{d-diff}, we fix the relative normalization by (see \re{app:Delta})
\begin{align}\label{applications-normalization}\notag
\Delta=d_--d_+{}&=\int_{-1}^1\frac{dt}{t}\,\rchi(x(t))\,x(t)
\\
{}& =\frac{\pi}{2}\int_{-\infty}^\infty
\frac{x\rchi(x)}{\sinh(\pi x/2)\cosh(\pi x/2)}\,dx\, .
\end{align}
With this choice, the determinant ratio is
\begin{equation}\label{applications-ratio}
\frac{D_-(\lambda)}{D_+(\lambda)}=Q_+(0|\lambda)\, .
\end{equation} 
Together with \re{D-W0} and the parity factorization, this gives 
\begin{align}\label{eq:Dpm}\notag
&\partial_\lambda\log D(\lambda)=d+f_\infty(\lambda)-f_\infty(0)
+\sum_{a}\,\bigl[f_a(\lambda)-f_a(0)\bigr]\,,
\\
&\partial_\lambda\log D_\pm(\lambda)=\frac12\left(\partial_\lambda\log D(\lambda)\mp\partial_\lambda\log Q_+(0|\lambda)\right)\,,
\end{align} 
where the functions $f_\infty(\lambda)$ and $f_a(\lambda)$ are given by 
\re{W-infty} and \re{fa}, respectively.

Parity simplifies the potentials \re{rational-symmetric-potentials} to
\begin{equation}\label{applications-potentials}
V_{++}(\lambda)=-\frac{4i}{\pi^2\lambda}Q_+'(i|\lambda)\, ,\qqqquad
V_{--}(\lambda)=\frac4{\pi^2\lambda}\bigl[1-Q_-'(i|\lambda)\bigr]\, .
\end{equation}  
Substituting these relations into \re{W-infty}, we obtain
\begin{align}\notag\label{ff's}
    &f_\infty(\lambda)=2c_\infty\bigl[Q_-'(i|\lambda)+iQ_+'(i|\lambda)-1\bigr]-\frac{2i d_\infty}{\pi \lambda} Q_+'(i|\lambda) \,,
 \\
 &f_\infty(0)=-\frac{2i d_\infty}{\pi}
 \partial_\lambda Q_+'(i|0)= d_\infty \int_{-1}^1dt\,\rchi(x(t))\, .
\end{align}
Using $Q_+(x|0)=1$ and $Q_-(x|0)=x$, the finite-pole contribution \re{eq:finite-pole} becomes  
\begin{align}\label{applications-sector-difference}\notag
& f_a(\lambda)=\frac1{2(1-t_a^2)}\bigg[
\left(c_a-\frac{d_a}{\pi\lambda}\right)\Wr(Q_+^-,Q_-^-)
+c_a\bigl(4-\Wr(Q_+^+,Q_-^+)\bigr)\bigg] \, ,
\\[1.5mm]
&f_a(0)=d_a \int_{-1}^1 \frac{dt\,\rchi(x(t))}{(t_a-t)^2}\, ,
\end{align} 
with the shift combinations $Q_\sigma^\pm$ and the differential Wronskian defined as in \re{Q-eps} and \re{Wron}, with all functions evaluated at $x_a$ \re{finite-pole-points}. The second relation in \re{applications-sector-difference} follows from \re{fa}.

These expressions determine the logarithmic derivative of the full determinant through the first relation in \re{eq:Dpm}. To recover the individual sector derivatives from the second relation in \re{eq:Dpm}, we also need the logarithmic derivative of $Q_+(0|\lambda)$. It admits the localization formula
\begin{align}
	\label{sector-difference}
	{}&\partial_\lambda\log Q_+(0|\lambda)=
	2c_\infty Q_-'(i|\lambda)
	-2i\left(c_\infty-\frac{d_\infty}{\pi\lambda}\right)Q_+'(i|\lambda)\notag
    \\
	{}&\qqqquad\qquad+\frac{\pi}{4}\sum_a\frac1{t_a}
	\left[c_a Q_+^+(x_a|\lambda)Q_-^+(x_a|\lambda)
	-\left(c_a-\frac{d_a}{\pi\lambda}\right)Q_+^-(x_a|\lambda)Q_-^-(x_a|\lambda)\right],
\end{align}  
where the Baxter $Q-$functions are evaluated at $x=i$ and $x=x_a$, corresponding to the poles of the functions \re{AB} at $t\to\infty$ and $t=t_a$, respectively.

\subsection*{Equal-mass QCD$_2$}\label{combined-QCD-example}
For equal masses, with \(\beta=0\) and \(\alpha>-1\), the symbol function is given by \re{chi-gen} and \re{AB} with
\begin{equation}
c_\infty=1\, ,\qquad
d_\infty=\frac{2\alpha}{\pi}\, ,\qquad
c_a=d_a=0\, .
\end{equation}
The relative normalization obeys $d_- - d_+=\Delta_{\rm QCD}=2-\alpha V_{++}(0)$ (see \re{d-diff} and the equal-mass limit of \re{V''}). Moreover, the regularization constant $d_-$ is obtained from \re{d-QCD} by taking the limit $\beta\to0$,
\begin{align}
d_-=2+\frac{\alpha}{4}i_2(\alpha)\, .
\end{align}
Substituting there results into \re{eq:Dpm} -- \re{sector-difference}, we find that
\begin{align}\label{qcd-sum-difference}
&\partial_\lambda\log D(\lambda)=2Q_-'(i|\lambda)
+2i\left(1-\frac{2\alpha}{\pi^2\lambda}\right)Q_+'(i|\lambda)
+\frac{\alpha}{2}i_2(\alpha)\, ,
\notag\\[2mm]
&\partial_\lambda\log Q_+(0|\lambda)=2Q_-'(i|\lambda)
-2i\left(1-\frac{2\alpha}{\pi^2\lambda}\right)Q_+'(i|\lambda)\, .
\end{align}
Combining these relations according to \re{eq:Dpm} yields the sector identities
\begin{align}\label{spec-QCD}
\partial_\lambda\log D_-(\lambda)
&=2Q_-'(i|\lambda)
+\frac{\alpha}{4}i_2(\alpha)\, ,
\notag\\[2mm]
\partial_\lambda\log D_+(\lambda)
&=2i\left(1-\frac{2\alpha}{\pi^2\lambda}\right)Q_+'(i|\lambda)
+\frac{\alpha}{4}i_2(\alpha)\, .
\end{align}
Using \(Q_+(i|\lambda)=1\) and \(Q_-(i|\lambda)=i\), these expressions reproduce \re{localization-equal-masses} and the conjectures of \cite{Fateev:2009jf,Litvinov:2024riz}.

\subsection*{Ising field theory}
For the Ising symbol \re{chi-Ising}, with $\beta=0$ and $\alpha>0$, the functions $A(t)$ and $B(t)$ in (\ref{chi-gen}) are proportional to each other and take the following form
\begin{equation}\label{ising-rational-closure}
A(t)=\frac1t\, ,\qquad B(t)=\frac{2\alpha}{\pi t}\, .
\end{equation} 
We observe that both functions exhibit a pole  on the integration interval $[-1,1]$ at $t=0$.  We can circumvent this problem by introducing 
\begin{equation}\label{ising-pole-regularization}
A_\epsilon(t)=\frac{t}{t^2+\epsilon^2}\, ,\qquad
B_\epsilon(t)=\frac{2\alpha}{\pi}A_\epsilon(t)\, ,\qquad \epsilon>0\, ,
\end{equation} which implement a principal-value prescription. 

At finite $\epsilon$, the localization identities are applied to $\rchi_\epsilon(x)=A_\epsilon(t)/(x+B_\epsilon(t))$, which reduces to the Ising symbol (\ref{chi-Ising})  as $\epsilon\to0$. 
The two poles $t_{1,2}=\pm i\epsilon$ have $c_1=c_2=-1/2$ and $d_1=d_2=-\alpha/\pi$, while $c_\infty=d_\infty=0$. As a consequence, $f_\infty=0$, while \re{applications-sector-difference} gives
\begin{equation}\label{ising-contact-subtraction}
\lim_{\epsilon\to0}\,\bigl[f_1(0)+f_2(0)\bigr]
={4\alpha\rchi(0)\over \pi}-\frac{2\alpha}{\pi}\int_{-1}^1
\frac{\rchi(x(t))-\rchi(0)}{t^2}\,dt
=2+\Delta_{\rm Ising}\, ,
\end{equation}
where $\rchi(0)=\pi/(2\alpha)$ and the relative normalization $\Delta_{\rm Ising}$ is defined in \re{applications-normalization}. We observe that $\rchi(x(t))-\rchi(0)=O(t^2)$, ensuring that the integral is convergent. 

Applying \re{eq:Dpm} and \re{sector-difference} and using the Baxter and Wronskian relations \footnote{The Baxter equation \re{Q-TQ} and its derivative at $x=0$ give, by parity, $Q_+(2i)=(1-\pi^2\lambda/\alpha)Q_+(0)$ and $Q_-'(2i)=(1-\pi^2\lambda/\alpha)Q_-'(0)$. The discrete Wronskian \re{seed-Q-Wronskian} at $z=0$ gives $Q_+(0)Q_-(2i)=2i$.}, we obtain 
\begin{align}\label{ising-sum-difference}
&\partial_\lambda\log D(\lambda)
=2i\left[
\left(1-\frac{\alpha}{\pi^2\lambda}\right)( \log Q_+(2i))'
+\frac{2\alpha-\pi^2\lambda}{\alpha-\pi^2\lambda}( \log Q_-(2i))'
\right] +2d_+-4\, ,
\notag\\[2mm]
&\partial_\lambda\log Q_+(0|\lambda)
=2i\left[
\left(1-\frac{\alpha}{\pi^2\lambda}\right)(\log Q_+(2i))'
-\frac{2\alpha-\pi^2\lambda}{\alpha-\pi^2\lambda}(\log Q_-(2i))'
\right]  \, .
\end{align}
The Baxter functions in this relation depend on the spectral parameter $\lambda$, the derivatives act on their first argument $x$ and the values at $x=2i$ are reached from within the fundamental strip. Combining \re{ising-sum-difference} with \re{eq:Dpm}, we find 
\begin{align}\label{applications-Ising-localization}
\partial_\lambda\log D_-(\lambda)
&=2i\left(1-\frac{\alpha}{\pi^2\lambda}\right)
(\log Q_+(2i))'+d_+-2\, ,
\notag\\[2mm]
\partial_\lambda\log D_+(\lambda)
&=2i \lr{1+\frac{1}{1-\frac{\pi ^2 \lambda }{\alpha }}}
(\log Q_-(2i))'+d_+-2\, .
\end{align}
The subtraction constant $d_+$ depends on the regularization scheme. For $d_+=2+\alpha i_2(\alpha)/4$ these expressions reproduce the conjecture of Ref.~\cite{Litvinov:2025geb}.

\paragraph{Physical odd-sector determinant.}
The identities above refer to the common integral problem. The physical Ising Bethe--Salpeter equation contains an additional odd rank-one term \cite{Litvinov:2025geb},
\begin{equation}\label{example-Ising-metric}
\bm\rho_{\rm phys}(\lambda,0)
=\bm\rho(\lambda,0)-\frac1{16}\ket{e_-}\bra{e_-}\, .
\end{equation}
Keeping the subtraction convention inherited from the auxiliary determinant, (\ref{example-Ising-metric}) implies 
\begin{equation}\label{applications-Ising-physical}
D_-^{\rm phys}(\lambda)
=D_-(\lambda)\frac{1-V_{--}(\lambda)/16}{1-V_{--}(0)/16}\, ,
\end{equation}
where $V_{--}$ is given by \re{applications-potentials}, while the even-sector determinant is unchanged. It follows from the previous expression that $\partial_\lambda \log D^{\rm phys}_-(\lambda)$ acquires the additional term $\partial_\lambda\log(1-V_{--}(\lambda)/16)$. In particular, this convention fixes
\begin{equation}\label{example-Ising-physical-slope}
d_-^{\rm phys}
=d_--\frac{\left.\partial_\lambda V_{--}(\lambda)\right|_{\lambda=0}}
{16-V_{--}(0)}\, .
\end{equation}

\subsection*{A quartic gauge-field potential}
This theory is obtained from QCD$_2$ by replacing the quadratic gauge-field potential with a quartic polynomial~\cite{Ambrosino:2024prz},
\begin{equation}\label{sec5-generalized-potential}
V(B)=\frac{B^2}{8\pi}+\frac{g_4B^4}{32\pi^3}\, .
\end{equation}
For equal quark masses, the meson spectrum is described by \re{rho} with $\beta=0$ and \footnote{Here $x$ is the variable denoted by $\nu$ in \cite{Ambrosino:2024prz}, and the normalisation of $\lambda$ is unchanged.}
\begin{equation}\label{example-generalized-weight}
\rchi(x)=\bigg[
x\left(\frac{1-g_4}{t}+\frac{2g_4}{t^3}\right)
+\frac2\pi\left(\alpha+\frac{g_4}{3}-\frac{2g_4}{t^2}\right)
\bigg]^{-1}\, .
\end{equation}
For $\alpha>-1$ and $g_4>0$, this symbol is positive and regular on the real axis, with $\rchi(0)=\pi/[2(\alpha+1)]$. Matching to \re{chi-gen} gives
\begin{equation}\label{quartic-rational-closure}
A(t)=\frac{t^3}{(1-g_4)t^2+2g_4}\, ,\qquad
B(t)=\frac{2t[(\alpha+g_4/3)t^2-2g_4]}{\pi[(1-g_4)t^2+2g_4]}\, .
\end{equation}
For $g_4\ne1$, the poles and coefficients in \re{AB} are
\begin{align}\label{example-generalized-coefficients}\notag
& t_{1,2}=\pm\sqrt{\frac{2g_4}{g_4-1}}\,,  \qquad
  c_\infty=\frac1{1-g_4}\, ,
\qquad d_\infty=\frac{2(\alpha+g_4/3)}{\pi(1-g_4)}\, ,
\\[2mm]
& c_1=c_2=\frac{g_4}{(1-g_4)^2}\, , \qquad
 d_1=d_2 =\frac{2g_4(\alpha+1-2g_4/3)}{\pi(1-g_4)^2}\, .
\end{align}
The poles are imaginary for $0<g_4<1$ and real outside $[-1,1]$ for $g_4>1$. Choosing $t_1$ on the positive imaginary or positive real axis, respectively, we find that $x_1$ (\ref{finite-pole-points}) is given by 
\begin{equation}\label{quartic-evaluation-point}
x_1=
\begin{cases}
\displaystyle\frac{2i}{\pi}\arctan\sqrt{\frac{2g_4}{1-g_4}}\, ,&0<g_4<1\, ,\\[3mm]
\displaystyle\frac1\pi\log\frac{t_1+1}{t_1-1}+i\, ,&g_4>1\, .
\end{cases}
\end{equation}
Since $f_2(\lambda)=f_1(\lambda)$,  the localization points are $x=i$, $x=x_1$ and $x=x_1-2i$.
The contribution of the pole $t_1$ follows from \re{applications-sector-difference}
\begin{align}\label{applications-quartic-contraction}
f_1(\lambda)
&=\frac{g_4}{2(1-g_4^2)}\Bigg[
4-\Wr(Q_+^+,Q_-^+)
+\left(1-\frac{2(\alpha+1-2g_4/3)}{\pi^2\lambda}\right)
\Wr(Q_+^-,Q_-^-)
\Bigg]_{x=x_1}\, ,
\end{align}
where the shift combinations and $\Wr$ are defined in \re{Q-eps} and \re{Wron}, and all Baxter functions have spectral parameter $\lambda$. The pole subtraction $f_1(0)$ is given by \re{applications-sector-difference}.

Combining these results, we find that \re{D-W0} and \re{sector-difference} can be written as follows
\begin{align}\label{D-W1}
&\partial_\lambda\log D(\lambda)
=d+\frac{2}{1-g_4}\bigl[Q_-'(i|\lambda)-1+iQ_+'(i|\lambda)\bigr]
+2\bigl[f_1(\lambda)-f_1(0)\bigr]\notag\\[1mm]
&\hspace*{26mm}-\frac{4i(\alpha+g_4/3)}{\pi^2(1-g_4)}
\left[\frac{Q_+'(i|\lambda)}{\lambda}
-\left.\partial_\lambda Q_+'(i|\lambda)\right|_{\lambda=0}\right]\, ,
\\[2mm]\label{quartic-sector-difference}
&\partial_\lambda\log Q_+(0|\lambda)
=\frac{2}{1-g_4}\left[
Q_-'(i|\lambda)
-i\left(1-\frac{2(\alpha+g_4/3)}{\pi^2\lambda}\right)Q_+'(i|\lambda)
\right]\notag\\[1.2mm]
&\hspace*{26mm}+\frac{\pi g_4}{2(1-g_4)^2t_1}\bigg[
Q_+^+Q_-^+
-\left(1-\frac{2(\alpha+1-2g_4/3)}{\pi^2\lambda}\right)
Q_+^-Q_-^-
\bigg]_{x=x_1}\, .
\end{align}
Combining \re{D-W1} and \re{quartic-sector-difference} via \re{eq:Dpm} yields the two sector identities at the same three localization points. 

\subsection*{Mixed symbols}

As a final example, consider a symbol combining the QCD and Ising terms with an additional rational contribution,
\begin{equation}\label{applications-mixed-symbol}
\rchi(x)=\bigg[
\frac{2\alpha}{\pi}
+x\left(\frac{a_1}{t}+a_2t\right)
+\frac{b}{\tau^2-t^2}
\bigg]^{-1}\, ,
\end{equation}
where $a_1,a_2>0$, $\alpha,b\ge0$ and $\tau>1$. The symbol is positive and regular on the real axis, with $\rchi(0)=[2(\alpha+a_1)/\pi+b/\tau^2]^{-1}$. Matching the previous expression to \re{chi-gen} gives
\begin{equation}\label{mixed-rational-closure}
A(t)=\frac{t}{a_1+a_2t^2}\, ,\qquad
B(t)=\frac{t}{a_1+a_2t^2}\left(\frac{2\alpha}{\pi}+\frac{b}{\tau^2-t^2}\right)\, .
\end{equation}
With $\kappa=\sqrt{a_1/a_2}$, the poles and coefficients in \re{AB} are
\begin{equation}\label{example-multipole-coefficients}
\begin{aligned}
t_{1,2}&=\pm i\kappa\, ,& \qquad
&c_1=c_2=-\frac1{2a_2}\, ,& \qquad
d_1=d_2&=-\frac1{2a_2}
\left(\frac{2\alpha}{\pi}+\frac{b}{\tau^2+\kappa^2}\right)\, ,\\[2mm]
t_{3,4}&=\pm\tau\, ,& &
c_3=c_4=0\, ,&
d_3=d_4&=\frac{b}{2(a_1+a_2\tau^2)}\, ,\\[2mm]
c_\infty&=0\, ,& &d_\infty=0\, .
\end{aligned}
\end{equation}
The imaginary poles occur in $A(t)$, whereas the real poles occur only in $B(t)$ and disappear when $b=0$. Choosing $t_1=i\kappa$ and $t_3=\tau$, we find the points specified by \re{finite-pole-points},
\begin{equation}\label{example-multipole-points}
x_1=\frac{2i}{\pi}\arctan\kappa\, ,\qquad
x_3=\frac1\pi\log\frac{\tau+1}{\tau-1}+i\, .
\end{equation}
Since $f_2(\lambda)=f_1(\lambda)$ and $f_4(\lambda)=f_3(\lambda)$, the localization points are $x_1$ and $x_1-2i$, together with $x_3$ and $x_3-2i$ when $b\ne0$. The contributions of the poles $t_1$ and $t_3$ follow from \re{applications-sector-difference},
\begin{align}\label{applications-mixed-contraction}
f_1(\lambda)
&=-\frac1{4(a_1+a_2)}\bigg[
4-\Wr(Q_+^+,Q_-^+)
+\left(1-\frac{2\alpha/\pi+b/(\tau^2+\kappa^2)}{\pi\lambda}\right)
\Wr(Q_+^-,Q_-^-)
\bigg]_{x=x_1}\, ,\notag\\[2mm]
f_3(\lambda)
&=\frac{b}
{4\pi\lambda(a_1+a_2\tau^2)(\tau^2-1)}
\left.\Wr(Q_+^-,Q_-^-)\right|_{x=x_3}\, ,
\end{align}
where the pole subtractions $f_1(0)$ and $f_3(0)$ are given by \re{applications-sector-difference}. Combining these results, we find that \re{D-W0} and \re{sector-difference} take the form
\begin{align}\label{mixed-sum}
&\partial_\lambda\log D(\lambda)
=d+2\bigl[f_1(\lambda)-f_1(0)\bigr]
+2\bigl[f_3(\lambda)-f_3(0)\bigr]\, ,
\\[2mm]\label{mixed-sector-difference}
&\partial_\lambda\log Q_+(0|\lambda)
=\frac{\pi}{4a_2t_1}\bigg[
\left(1-\frac{2\alpha/\pi+b/(\tau^2+\kappa^2)}{\pi\lambda}\right)
Q_+^-Q_-^-
-Q_+^+Q_-^+
\bigg]_{x=x_1}
\notag \\ 
&\hspace*{25mm}
+\frac{b}{4\lambda\tau(a_1+a_2\tau^2)}
\left[Q_+^-Q_-^-\right]_{x=x_3}\, .
\end{align}
Combining \re{mixed-sum} and \re{mixed-sector-difference} via \re{eq:Dpm} yields the two sector identities at the same localization points.

For $b=0$, $a_1=1$ and $a_2\to0$, one has $x_1\to i$ and recovers the QCD$_2$ identities with the corresponding subtraction convention. For $b=0$, $a_2=1$ and $a_1\to0$ with $\alpha>0$, one recovers the Ising case. Here $\epsilon^2=a_1$ in \re{ising-pole-regularization}, so the combined limit retains the constant $2$ in \re{ising-contact-subtraction}.

\section{Localization for general kernels}
\label{sect6}

In the previous section, we extended the localization identities to rational symbols \re{chi-gen} and \re{AB} while keeping the kernels \re{HK} fixed.
We demonstrated that using $t=\tanh(\pi x/2)$ as the independent variable significantly simplifies the analysis: 
the real $x$ axis is mapped to the interval $t\in [-1,1]$, the operator $\bm H$ becomes a Cauchy operator supported on a single cut $\Sigma=(-1,1)$, and
the reference state  $e_+(x)=\vev{x|e_+}$ becomes constant.

We now address the remaining
question: which changes of the reference state $\ket{e_+}$ and support $\Sigma$ preserve the localization identities for the determinant of the spectral operator? 
We formulate the kernel data as a seed and derive the general expression for the operators satisfying \re{KH-algebra}. This identifies
the additional condition needed for finite-rank virial closure.
The general form of the spectral operator \re{rho} and resolvent conventions remain those of the preceding
sections. 

\subsection{Seed construction}

\label{sec:seed-ingredients}

A seed $(\bm t, e_+,\bm H)$ consists of the Hermitian operator $\bm t$ acting by multiplication by $t$ on $L^2(\Sigma,dt)$,
a real, nonvanishing source $e_+(t)=\vev{t|e_+}$, and an anti-Hermitian operator $\bm H$ satisfying
\begin{equation}\label{seed-commutator}
 [\bm t,\bm H]=\ket{e_+}\bra{e_+},
 \qquad \bm H^\dagger=-\bm H.
\end{equation}
We assume that the support $\Sigma$ is symmetric, $e_+(t)$ is even, and the operator
$\bm H$ is $\bm{CP}$-odd, with $\bm C$ and $\bm P$ defined as in
Section~\ref{sec:KH-properties}. The symbol $\rchi(x)$ is assumed to be a real even function. Together with the couplings $\lambda$ and $\beta$ it
specifies the spectral operator \re{rho}.

In the $t-$representation, the commutator \re{seed-commutator}
fixes the kernel $H(t,t')=\vev{t|\bm H|t'}$ as~\footnote{The relation \re{seed-commutator} allows one to add to $H(t,t')$ an additional
multiplication operator $i c(t)\delta(t-t')$. Anti-Hermiticity of $\bm H$ requires real $c(t)$. We set $c(t)=0$ throughout. }
\begin{equation}\label{seed-kernel-freedom}
 H(t,t')=e_+(t)e_+(t')\,\mathrm{PV}\frac1{t-t'}\,.
\end{equation}
where $\mathrm{PV}$ denotes the Cauchy principal value. 
The multiplication operator $\bm x=x(\bm t)$
and the integrable operator $\bm K$ are defined through
\begin{equation}\label{seed-VK}
 \bm H\ket{e_+}=2\bm x\ket{e_+},\qqqquad
 \bm K\equiv\frac\pi2[\bm x,\bm H].
\end{equation}
Integrating the kernel \re{seed-kernel-freedom} against the reference state $e_+(t')$  we use the first relation to find
\begin{equation}\label{seed-V-explicit}
 x(t)\equiv\frac12\fint_\Sigma\frac{e^2_+(t')}{t-t'}\,dt'.
\end{equation}
Since $\Sigma$ is symmetric under $t\to -t$, it follows from this relation that the function $x(t)$ is real and odd. For $t \notin\Sigma$, the Cauchy transform
\re{seed-V-explicit} defines $x(t)$ as a function with cuts along $\Sigma$. For $t\in\Sigma$, the function inverse to $x(t)$
may have different branches. For instance, for $\Sigma$ being a union of two cuts, each cut maps onto the real
$x$ axis, giving two inverse branches $t_\pm(x)$.  

In a similar manner, the second relation in \re{seed-VK} gives the kernel $K(t,t')=\vev{t|\bm K|t'}$
\begin{equation}\label{seed-K-kernel}
 K(t,t')=\frac\pi2 e_+(t)\,
 \frac{x(t)-x(t')}{t-t'}\,e_+(t').
\end{equation}
This kernel has the integrable form \re{intro:integrable}. Its diagonal $K(t,t)$ is a finite function of $t$ for smooth $x(t)$.
  
To connect with the discussion in the previous sections, we consider the simplest case of a
single interval $\Sigma=(-1,1)$ and a constant source
$e_+(t)=\sqrt{2/\pi}$. With this choice, the relation \re{seed-V-explicit}
gives $x(t)$ given by \re{x(t)} and satisfying $t=\tanh(\pi x/2)$. The corresponding expressions for the kernels \re{seed-kernel-freedom} and \re{seed-K-kernel} are
\begin{align}\label{seed-concrete-VK} 
 H(t,t'){}&=\frac2\pi\,\mathrm{PV}\frac1{t-t'}\,,
\qqqquad
 K(t,t')=\frac{x(t)-x(t')}{t-t'}\,.
\end{align}
This relation defines the kernel of the operators $\bm H$ and $\bm K$ in the $t-$representation. Changing the variable to $x=x(t)$, it is straightforward to verify that in the $x-$representation the same kernels coincide with those in \re{HK}.
  
For a general source $e_+(t)$, we can apply \re{seed-commutator} and \re{seed-VK} to derive the counterparts of
\re{H-com} and \re{KH-algebra}. This allows us to identify the operator relations which are insensitive to the choice of
the source and support.
Since $\bm t$ and $\bm x$ commute, the Jacobi identity gives 
\begin{equation}\label{seed-K-commutator}
\frac{2}{\pi}[\bm t,\bm K]=[\bm x,[\bm t,\bm H]]= \bm x\ket{e_+}\bra{e_+}
                   -\ket{e_+}\bra{e_+}\bm x\,.
\end{equation}
Thus, the commutator $[\bm t,\bm K]$ has rank at most two.

The relation \re{seed-K-commutator} can be generalized for $\bm t$ replaced by a rational function $A(\bm t)$ of the form \re{AB}. The derivation of \re{fivebis:pole-commutator} uses only the rank-one identity \re{seed-commutator} and, therefore, it holds for every seed. Thus, the commutator $[A(\bm t),\bm H]$ has finite rank
 with poles outside $\Sigma$. Explicitly,
\begin{equation}\label{seed-rational-H}
 \vev{t|[A(\bm t),\bm H]|t'}
 =e_+(t)\frac{A(t)-A(t')}{t-t'}e_+(t').
\end{equation}
Since $A(\bm t)$ also commutes with $\bm x$, the Jacobi
identity gives
\begin{equation}\label{seed-rational-K}
 [A(\bm t),\bm K]
 =\frac\pi2[\bm x,[A(\bm t),\bm H]],
\end{equation}
which also has finite rank.

The seed also relates $\bm H^2$ to $\bm K$. Taking the
Hermitian conjugate of the first relation in \re{seed-VK} gives
$\bra{e_+}\bm H=-2\bra{e_+}\bm x$. Hence
\begin{equation}\label{seed-H-square-commutator}
 [\bm t,\bm H^2]
 =\bm H\ket{e_+}\bra{e_+}\,+\,\ket{e_+}\bra{e_+}\bm H
 \,=\,2\Big(\bm x\ket{e_+}\bra{e_+}\,-\,\ket{e_+}\bra{e_+}\bm x\Big)
 \,=\,\frac4\pi[\bm t,\bm K].
\end{equation}
It follows from this relation that $\bm H^2-(4/\pi)\bm K$ commutes with $\bm t$
and hence acts by multiplication 
\begin{equation}\label{seed-U}
 \bm H^2-\frac4\pi\bm K=U(\bm t).
\end{equation}
Substituting the explicit kernels \re{seed-kernel-freedom} and
\re{seed-K-kernel} into \re{seed-U} gives
\begin{equation}\label{seed-U-explicit}
 U(t)=-\pi^2e^4_+(t)\,.
\end{equation}
The off-diagonal part of $\bm H^2$ cancels $(4/\pi)K(t,t')$
in \re{seed-U}, leaving the local term
$-\pi^2e^4_+(t)\delta(t-t')$. Applying \re{seed-U} to the
source and using \re{seed-VK} then yields
\begin{equation}\label{seed-source-actions}
 \bm H\bm x\ket{e_+}
 =\left(\bm x^2+\tfrac14 U(\bm t)\right)\ket{e_+},\qquad
 \bm K\ket{e_+}=\frac\pi2\left(\bm x^2-\frac{1}{4}U(\bm t)\right)\ket{e_+}.
\end{equation}
For the constant-source seed $e_+(t)=\sqrt{2/\pi}$ we find from \re{seed-U-explicit}
that $U=-4$ and the relations \re{seed-source-actions} reduce to \re{c-rel}.
 
\subsection{The spectral problem}
\label{sec:seed-spectral}
  
To specify the spectral operator \re{rho}, we need only two additional ingredients:
a positive, even symbol $\rchi(t)$ and a real coupling $\beta$. As before, we assume that $\bm M=\bm\rchi^{-1}-i\beta\bm H/\pi>0$ which leads to additional conditions for
$\beta$ and the symbol $\rchi$. We also assume that a complete discrete basis of generalized eigenstates \re{eig1} 
is normalized as in \re{inner}. The completeness expansion \re{dec} and the $CP$ phase convention
in \re{psi-e} apply unchanged. These are assumptions on the spectral
problem, not consequences of the seed commutator.
 
We now extend the  sector decomposition \re{sec} and the virial identity \re{rational-norm-identity} to
a general seed construction. The derivation of the selection rule \re{vir-eq1} uses only the
rank-two commutator \re{seed-K-commutator} and the $CP$ phase convention \re{psi-e}. Therefore, it carries over for the seed with
$\ket{e_-}=\bm x\ket{e_+}$. We therefore retain the two continued spectral
sequences and determinant definitions of \re{sec} and
\re{Dpm}. For a general seed, the even and odd indices label
these sequences without implying that their levels interlace. Under the same spectral assumptions as before, the spectral determinants
are defined by \re{D-def} and \re{Dpm}.

The general expression for the symbol function \re{chi-gen} extends directly to the seed multiplier $x(t)$ defined in \re{seed-V-explicit}
\begin{equation}\label{seed-rational-weight}
 A(t)\rchi^{-1}(t) =x(t)+B(t).
\end{equation}
Recall that $x(t)$ is generally nonrational (see \re{x(t)}). We take $A(t)$ and $B(t)$ of the form
\re{AB}, with a linear term and simple poles $t_a$ outside $\Sigma$. Finite-rank
closure also requires $U(t)A(t)$ to be rational with poles
outside $\Sigma$. This restricts the source through
$U(t)=-\pi^2e^4_+(t)$ and holds automatically for constant sources.

In close analogy with  \re{fivebis:generator}, we choose $\bm q=iA(\bm t)\bm H$ and examine the virial identity \re{vir}. 
As before, the $\beta$ dependent terms cancel, the term depending on the symbol becomes $(2/\pi)\bm K+[B(\bm t),\bm H]$, and the relation \re{seed-U} gives
\begin{align} \label{seed-virial-operator}\notag
{}& \bm\rho A(\bm t)\bm H
 -\bm H A(\bm t)\bm\rho 
 =\frac2\pi\bm K+[B(\bm t),\bm H]-\frac{\pi\lambda}{4}\bm{\mathcal B},
 \\
{}&  \bm{\mathcal B} \equiv-\bm H[A(\bm t),\bm H]\bm H
                   -[U(\bm t)A(\bm t),\bm H].
\end{align}
Taking the expectation value of \re{seed-virial-operator} at
$\lambda=\lambda_n$, and using the normalization
\re{inner}, yields
\begin{equation}\label{seed-level-virial}
 \frac2{\pi\lambda_n}
 =\frac{\pi\lambda_n}{4}\vev{\psi_n|\bm{\mathcal B}|\psi_n}
              -\vev{\psi_n|[B(\bm t),\bm H]|\psi_n}.
\end{equation}
The operators inside the matrix elements on the right-hand side have finite rank by
\re{seed-rational-H} and \re{seed-virial-operator}. This extends the finite-rank reduction underlying
\re{rational-norm-identity} to sources satisfying the
additional rationality condition stated above.

The usual mechanism then applies. Inserting \re{seed-level-virial} into the spectral sum \re{D-W0} expresses $\partial_\lambda\log D_\pm$ through finitely many resolvent matrix elements between the source vectors. Once these matrix elements are identified with values of the Baxter functions at special points, this yields localization identities for a general seed. In full generality, however, this analysis becomes cumbersome, since both the source vectors and the evaluation points depend on the seed. In the next subsection we therefore illustrate the mechanism with a simple example.
 
\subsection{Recovery of the determinant identities}
\label{sec:seed-ratio}

We consider the simplest deformation of the constant-source seed  $e_+(t)=\sqrt{2/\pi}$  and then identify the steps that hold for a general seed.

We keep the single interval $\Sigma=(-1,1)$ and deform the constant source,
\begin{equation}\label{seed-example}
 e_+^2(t)=\frac2\pi\lr{1+\varepsilon t^2},\qqquad \varepsilon<1\,.
\end{equation}
 The relation \re{seed-V-explicit} gives
\begin{equation}\label{seed-example-x}
 x(t)=(1+\varepsilon t^2)\,x_0(t)-\frac{2\varepsilon}{\pi}\,t\,,\qqquad x_0(t)=\frac1\pi\log\frac{1+t}{1-t}\,,
\end{equation}
where $x_0(t)$ is the function \re{x(t)} corresponding to the constant-source seed. According to \re{seed-U-explicit}, the function
$U(t)=-4(1+\varepsilon t^2)^2$ is a polynomial, so that the closure condition of Section~\ref{sec:seed-spectral} is satisfied for any $A(t)$ of the form \re{AB}.

Let us consider the simplest symbol $\chi=t/(x(t)+2\alpha t/\pi)$, corresponding to $A(t)=t$ and $B(t)=2\alpha t/\pi$. Below we focus on the key new features due to the modification of the source \eqref{seed-example} with a term dependent on $t$.

First, the function $U(t)A(t)$ is now a polynomial of degree five. A polynomial can be obtained as a limit of the rational functions \re{AB} with poles sent to infinity, and in this limit the pole sources \re{fivebis:sources} become the vectors $\bm t^j\ket{e_+}$. The virial relation \re{seed-level-virial} therefore acquires bilinears in the higher moments $\vev{e_+|\bm t^j|\psi_n}$ with $j\le4$, which are the analogues of the pole terms in \re{rational-norm-identity}. 

Second, continuing $x(t)$ across the cut $-1<t<1$ shifts it by $i\pi e_+^2(t)$ instead of $2i$. Since this shift depends on $t$, the natural spectral parameter is
\begin{equation}\label{seed-example-u}
 u=\frac{2x(t)}{\pi e_+^2(t)}\,.
\end{equation}
It jumps across the cut by $2i$. 
Thus suggests to rewrite the Baxter equation \re{Q-TQ} in terms of $u$, rather than $x$.
Replacing $x$ with $u$, we get the Baxter equation for $Q_\pm(u)$ in the usual form \re{Q-TQ}, with an extra multiplier $-U(t)/4$ in the r.h.s.. This new multiplier, now depends on $u$ through the multivalued inverse map $t(u)$. For small $\varepsilon$, the branch points of $t(u)$ form a short cut of length $\sim\sqrt\varepsilon$ through $u=\pm i$, which shrinks to a regular point as $\varepsilon\to0$.

Third, the localization point acquires a new local structure. The rank-two commutator \re{seed-K-commutator} reduces the matrix elements entering $\partial_\lambda\log D_\pm$ to the moments $\vev{e_+|\bm t^k\bm\rho^{-1}|e_\pm}$, that is, to the expansion of the Baxter functions at $t=\infty$. Localization therefore still occurs at $u=\pm i$. However, since $e_+^2$ grows as $t^2$, we have $u\mp i\sim t^{-3}$, so the functions $Q_\pm(u)$ have a branch point of order three at $u=\pm i$. The logarithmic derivatives $\partial_\lambda\log D_\pm$ are then given by the first few coefficients of the expansion of $Q_\pm(u)$ in powers of $(u\mp i)^{1/3}$, instead of first derivatives.

The determinant-ratio identity \re{second}, by contrast, is unchanged, because its derivation uses only the seed algebra; only the value of $\Delta$ changes. 
 
To conclude this section, the seed construction is our attempt to identify the universal structure behind the localization identities. We have shown that many of these relations follow from a simple operator algebra, generated by the rank-one Cauchy commutator \re{seed-commutator}. Combined with analyticity away from the real axis, this algebra leads to the Baxter equation, whose values at special points are in turn linked back to the spectral determinants.

\section{Concluding remarks and further directions}\label{sect7}

Spectral determinants encode the complete spectrum of the meson bound states of planar QCD$_2$ and the Ising field theory in entire functions of the spectral parameter. Their zeros determine the squared meson masses, with the physical Ising spectrum obtained after the odd-sector projection and the rank-one modification discussed above. The expansion of their logarithms near the origin generates regularized spectral sums, while their asymptotic behaviour constrains the highly excited spectrum.
 
In this paper, we have derived localization identities that give access to these spectral data in planar QCD$_2$  and the Ising field theory, proving previously conjectured relations. They express logarithmic derivatives and ratios of sector determinants through Baxter $Q$-functions and their derivatives at special points, avoiding integration of the resolvent trace along the full contour. The derivation relies on finite-rank commutation relations and virial identities. We have also identified a sufficient rational closure condition that extends localization to a broader class of symbols and determines the evaluation points. The seed construction isolates the operator algebra underlying this reduction and provides a starting point for further generalizations.

The resulting framework suggests several directions for further study: changes of the reference state and integration support,
limits in which rational poles approach the contour, applications to spectral asymptotics, deformation equations, and extensions to
other integrable kernels. 

\paragraph{Several intervals and matrix Baxter equations.} 
We assumed above that the support $\Sigma$ consists of a single interval $(-1,1)$. Keeping the source $e_+(t)$ constant, we can replace its support by the union of two intervals $\Sigma=(-1,-a)\cup(a,1)$, with $0<a<1$. The relation \re{seed-V-explicit} gives
\begin{equation}\label{example-two-cut-v}
 x(t)=\frac1\pi\log\left|
 \frac{(t+1)(t-a)}{(t+a)(t-1)}\right| .
\end{equation}
Each interval maps onto the real $x$ axis, producing two inverse branches $t=t(x)$. These branches provide two components of the dressed source functions and lead to a matrix version of the Baxter problem. The natural extension is to derive the corresponding matrix Wronskian and determinant identities, and to determine how their normalization depends on the endpoints of the support. For several constant-density intervals, the same construction suggests one component for each interval. Establishing the analytic conditions for localization in this setting would extend the scalar framework to coupled spectral problems.

\paragraph{Poles approaching the integration contour.}
\label{sec:examples-Ising}
A further question concerns limits in which poles of the rational multipliers \re{AB} approach the support. The Ising symbol provides a
concrete example: its pole at $t=0$ can be approached by the
symmetric pair $t=\pm i\epsilon$. The Cauchy sources \re{fivebis:sources} then acquire both principal-value and
$\delta$-function contributions. Their contractions must be evaluated
at finite $\epsilon$ before the limit is taken; averaging the
sources first can discard local contributions. A systematic
treatment should establish convergence of the regularized
determinants and identify which local terms survive in the
localization identities. This would clarify the boundary
prescription used for the Ising example and its extension to
higher-order poles or poles approaching endpoints of the support.

\paragraph{Spectral asymptotics and nonperturbative corrections.}
A direct application of the generalized localization identities is
to extract spectral information for the quartic deformation of
QCD$_2$ and the mixed symbols of Section~\ref{sec:applications}.
The expansion near $\lambda=0$ gives spectral sums, while the
large-$\lambda$ behaviour of the Baxter functions determines the
asymptotics of the determinants and, after the appropriate analytic
continuation, quantization conditions for their zeros. It would be
interesting to develop these expansions with explicit dependence
on the deformation parameters and to determine how the additional
localization points contribute to the highly excited spectrum.

The same analysis could address contributions beyond the asymptotic
power series. This requires identifying the relevant Baxter
solutions in different sectors of the complex spectral plane and
matching them across Stokes lines. The localization identities
would then translate the resulting connection data into
exponentially small corrections to the determinants and spectral
levels. Comparing these predictions with direct numerical spectra
would test the generalized formulas beyond the perturbative
expansion at the origin.

\paragraph{Deformation equations for the determinants.}
The coefficients and pole positions in the rational symbol provide
natural deformation parameters. Differentiating the resolvent with
respect to these parameters generates matrix elements with
insertions of the differentiated spectral operator. The finite-rank
commutator algebra suggests asking whether these insertions and the
original source contractions can be organized into a finite closed
system. Such closure would have to be established separately:
finite rank of the virial commutators does not by itself imply
closure under parameter differentiation.

If a closed system exists, it could give differential equations
for the spectral determinants as functions of the deformation
parameters, with boundary data fixed by the QCD or Ising limits.
The relation between these equations and the Baxter equation would
then provide a concrete setting in which to investigate the
connection with isomonodromic deformations and tau functions
mentioned in the Introduction. The mixed symbols offer a useful
test because their parameters move the localization points and
interpolate between distinct spectral problems.

\paragraph{Beyond rational closure.}
The condition \re{chi-gen} is sufficient for the finite-rank virial
reduction, but its necessity remains open. Within the present seed
construction, rational multipliers lead to finite sums of source
outer products through their divided differences. A useful first
step would be to determine whether this exhausts the admissible
multipliers under suitable regularity assumptions. A broader
question is whether alternative virial generators can produce
localization for symbols outside the class \re{chi-gen}.

This question also requires separating algebraic closure from
analytic localization. A finite set of source contractions need not
be expressible through values of Baxter functions at finitely many
points without additional information about their continuation.
A classification should therefore identify both the operator
relations responsible for finite rank and the analytic conditions
needed to evaluate the resulting contractions. This would clarify
which features of localization are specific to the kernels studied
here and which persist in a wider class of spectral problems.

\paragraph{Other integrable kernels.}
It would be natural to test the same mechanism for the ABJM
Fermi-gas and truncated Bessel operators discussed in the
Introduction. The first task is to identify multiplication
operators and reference states for which the relevant commutators
have finite rank, and then to seek a virial relation compatible
with the physical symbol. If these ingredients are available, one
can ask whether the resulting resolvent contractions admit an
evaluation at special points, analogous to the identities derived
here.
The appropriate continuation equations and determinant
normalizations would have to be determined for each kernel.
Establishing such examples would test whether the operator origin
of localization extends to the higher-dimensional observables that
motivate the Fredholm-determinant approach.
\paragraph{'t Hooft equations in curved backgrounds.}
The 't Hooft equation determines the meson spectrum of planar QCD with massive quarks in two-dimensional Minkowski spacetime. Extending the model to curved backgrounds would allow one to study how curvature modifies this spectrum. The case of (A)dS$_2$ was studied in \cite{Betzios:2026btz}, where it was found that the localization mechanism only applies partially: while the ratio of spectral determinants can still be related to special values of the corresponding Baxter $Q$-functions, the derivative relations no longer hold. It would be interesting to understand whether the operator approach developed here can be extended to this case. 

\section*{Note added}

During the final stages of preparing this paper, we learned that the localization identities in two-dimensional QCD had been independently derived in Ref.~\cite{Litvinov}. We thank the authors for sharing their draft with us and for coordinating the submission of our papers to arXiv.

\section*{Acknowledgements}

We would like to thank Federico Ambrosino, Alexander Gorsky, Vladimir Kazakov and Ivan Kostov for interesting discussions. The research
of B.B. was supported by the Doctoral Excellence Fellowship Programme funded by the National Research Development and Innovation Fund of the Ministry of Culture and Innovation and the Budapest
University of Technology and Economics, under a grant agreement with the National Research, Development and Innovation Office (NKFIH). It was also supported by the grant NKKP Advanced 152467. The work of G.K. and A.T. was supported by the French
National Agency for Research grant ``Observables'' (ANR-24-CE31-7996). The work of N.G. was supported by the Science and Technology Facilities Council (STFC) under the grant ST/X000753/1. N.G. is grateful to the Perimeter Institute and Yale University for their hospitality while part of this work was done. N.G. and G.K. are grateful to the organizers of the workshop ``Exact techniques and their applications -- inTropea 2026'' for creating a productive environment.
 
\section*{Use of AI tools}

AI tools were used in the preparation of this work. The derivations of the core examples were carried out independently by the authors and with the assistance of the top LLM models, and the results were cross-checked. Symbolic and numerical verifications were performed in Mathematica through the Wolfbook interface~\cite{Wolfbook}, which records the computations of both the authors and the AI in shared notebooks and revise latex notes collaboratively, so that each step can be inspected and validated. The authors take full responsibility for the content of the paper.

\appendix

\section{Normalization constants and Wronskian relation}\label{app}

In this appendix, we compute the small-$\lambda$ expansion of the Baxter function
$Q_+(x|\lambda)$ defined in \re{Q-defs}, to first order in $\lambda$. We then use it to fix the
normalization constants entering the localization identities and to derive the Wronskian
relation for the solutions of the Baxter equation.
 
\paragraph{Baxter function at small $\lambda$.}
We introduce $t=\tanh(\pi x/2)$ and define the $Q-$function at $\lambda=0$ as
\begin{align}\label{V-Cauchy-variable}
  q_+(t) = Q_+(x(t)|0) \,,\qqqquad
  x(t)={1\over\pi} \log\lr{1+t\over 1-t}\,.
\end{align}
The direct integral continuation of $q_+(t)$ to complex $t$ is given by $\tilde q_+(t)= \tilde Q_+(x(t)|0)$. The two functions are related to each other through \re{Q-Baxter-continuation}.
At $\lambda=0$, equations \re{Q-defs} and \re{Q-direct-continuation} reduce to the scalar Cauchy problem
\begin{align}\label{q-eq1} 
q_+(t)&=1+\frac{2i\beta}{\pi^2}\fint_{-1}^{1}
\frac{\rchi(x(t'))q_+(t')}{t-t'}\,dt', \qquad t\in[-1,1],
\\\label{V-Cauchy-integral}
\tilde q_+(t)&=1+\frac{2i\beta}{\pi^2}\int_{-1}^{1}
\frac{\rchi(x(t'))q_+(t')}{t-t'}\,dt',
\qquad t\notin[-1,1].
\end{align}
The boundary relation \re{Q-boundary} gives
\begin{align}\label{V-Cauchy-jump}
\frac{\tilde q_+(t+i0)}{\tilde q_+(t-i0)}=\frac{\Delta_+(x(t))}{\Delta_-(x(t))}\,,
\end{align}
where the dressing factors $\Delta_\pm(x)$ are defined in \re{Q-dressing-factors}.
In the parameter range $\alpha+1>|\beta|$, both dressing factors are positive for $t\in[-1,1]$. Taking the logarithm of their ratio, imposing $\tilde q_+(t)\to1$ as $t\to\infty$, and retaining the endpoint behaviour dictated by \re{V-Cauchy-integral}, we obtain
\begin{align}\label{q-int}
\tilde q_+(t)=\exp\left[\frac{1}{2\pi i}\int_{-1}^1\frac{dt'}{t'-t}\log\frac{\Delta_+(x(t'))}{\Delta_-(x(t'))}\right]\,.
\end{align}
For $-1\le t \le 1$  the matching relation \re{Q-boundary} gives
\begin{align}\label{q+}
q_+(t)= \left[{\tilde q_+(t+i0)\tilde q_+(t-i0)\over \Delta_+(x(t))\Delta_-(x(t))}\right]^{1/2}.
\end{align}
The square root is chosen continuously from $q_+(t)=1$ at $\beta=0$. Returning to the original variable yields, for real $x$,
\begin{equation}\label{app-Q-zero-real}
Q_+(x|0)=\frac{1}{\sqrt{\Delta_+(x)\Delta_-(x)}}\exp\left[\frac{1}{2\pi i}\fint_{-1}^1 \frac{dt'}{t'-t(x)}\log\frac{\Delta_+(x(t'))}{\Delta_-(x(t'))}\right],
\end{equation}
where $t(x)=\tanh(\pi x/2)$.

At $x=0$, the principal-value integral in the exponent of \re{app-Q-zero-real} vanishes because its integrand is odd. Hence
\begin{equation}\label{Q00}
Q_+(0|0)=\frac{1}{\sqrt{\Delta_+(0)\Delta_-(0)}}
=\frac{\alpha+1}{\gamma},
\end{equation}
where $\rchi(0)=\pi/[2(\alpha+1)]$ and $\gamma=\sqrt{(\alpha+1)^2-\beta^2}$. For real $x$, the exponent in \re{app-Q-zero-real} is imaginary and odd, implying
\begin{equation}\label{Q-sym}
\overline{Q_+(x|0)}=Q_+(-x|0)
=\frac{1}{\Delta_+(x)\Delta_-(x)Q_+(x|0)}.
\end{equation}

To obtain the derivative $\partial_\lambda\log Q_+(x(t)|\lambda)$ at $\lambda=0$, define  
\begin{align}
\dot q_+(t)\equiv \partial_\lambda  Q_+(x(t)|\lambda)\Big|_{\lambda=0}\,.
\end{align}
Differentiating \re{Q-defs} with respect to $\lambda$, setting $\lambda=0$ and combining the result with \re{q-eq1}, we find
\begin{align}
&x(t)q_+(t)-\frac{2i\beta}{\pi^2}\dot q_+(t)=x(t)+\frac{2i\beta}{\pi^2}\fint_{-1}^{1}
\frac{dt'}{t-t'}\rchi(x(t'))\left[x(t')q_+(t')-\frac{2i\beta}{\pi^2}\dot q_+(t')\right] .
\label{eq:source-x}
\end{align}
Notice that the combination $x(t)q_+(t)-\frac{2i\beta}{\pi^2}\dot q_+(t)$ satisfies the same scalar Cauchy equation as $q_+(t)$, with source $x(t)$ in place of $1$. Solving the corresponding inhomogeneous problem gives 
 \begin{align}
&x(t)q_+(t)-\frac{2i\beta}{\pi^2}\dot q_+(t)
\notag\\
&\quad=q_+(t)\bigg[
\frac{x(t)}
{\Delta_+(x(t))\Delta_-(x(t))q_+(t)}
 +
\frac{2i\beta}{\pi^2}\fint_{-1}^{1}
\frac{dt'}{t-t'}
\frac{\rchi(x(t'))\,x(t')}
{\Delta_+(x(t'))\Delta_-(x(t'))q_+(t')}\bigg].
\label{eq:inverse-x}
\end{align}
Using the scalar Cauchy solution \re{q-int} and \re{q+}, we simplify this expression to
\begin{align}\notag
\dot q_+(t){}&=q_+(t)\int_{-1}^{1}dt' {x(t)-x(t')\over t-t'}\frac{\rchi(x(t'))}{\Delta_+(x(t'))\Delta_-(x(t'))q_+(t')}
\\
{}&=q_+(t)\int_{-1}^{1}dt' {x(t)-x(t')\over t-t'}\rchi(x(t')) q_+(-t')\, ,
\end{align}
where the second equality follows from \re{Q-sym} and $x(-t')=-x(t')$.  
Evaluating at $t=0$ gives
\begin{align}\label{app:Delta} 
\Delta \equiv\partial_\lambda\log Q_+(0|\lambda)\Big|_{\lambda=0}{} =\int_{-1}^1\frac{dt'}{t'}\,\rchi(x(t'))\,x(t') q_+(t')\,,
\end{align}
where we used $\rchi(-x)=\rchi(x)$ and changed the integration variable $t'\to - t'$. 
For the QCD symbol \re{chi-QCD}, the identity $x(t)\rchi(x(t))/t=1-2\alpha\rchi(x(t))/\pi$ reduces \re{app:Delta} to
\begin{align}\label{Delta-QCD-app} 
\Delta_{\rm QCD}=\int_{-1}^1 dt' \left[1-\frac{2\alpha}{\pi}\rchi(x(t'))\right] q_+(t')\,.
\end{align}
Here the subscript is inserted to indicate that this relation holds specifically for the QCD${}_2$ symbol \re{chi-QCD}.
 
\paragraph{Constants $V_{++}(0)$ and $V'_{++}(0)$.}
Equations \re{V-Q}, \re{V-Q1} and \re{V-rel2} express these constants in terms of $q_+(t)$ as
\begin{align}\notag\label{app:V}
{}&V_{++}(0)=\frac{2}{\pi}\int_{-1}^{1}dt\,\rchi(x(t))q_+(t) \,,
\\
{}&V'_{++}(0) =\frac{\pi^2}{i\beta} V_{-+}(0)=\frac{2\pi}{i\beta}\int_{-1}^{1}dt\, x(t)\rchi(x(t))q_+(t) \,.
\end{align}
For $\beta\ne0$, the integral defining $V_{++}(0)$ determines the coefficient of $1/t$ in the large-$t$ expansion of \re{V-Cauchy-integral}
\begin{align} 
\tilde q_+(t)=1+{2i\beta\over \pi^2 t}\int_{-1}^1 dt'\,\rchi(x(t')) q_+(t') + O(1/t^2)\,.
\end{align} 
Matching this expansion to \re{q-int} yields
\begin{align}
V_{++}(0)={1\over 2\beta}\int_{-1}^1 dt\, \log {\Delta_+(x(t))\over \Delta_-(x(t))}\,.
\end{align}
For the QCD symbol, substituting the dressing factors \re{Q-dressing-factors} gives
\begin{align}\notag
{}&V_{++}(0)=\frac{(\alpha+\beta)i_2(\alpha+\beta)-(\alpha-\beta)i_2(\alpha-\beta)}{4\beta}\,,
\\
{}& i_2(\alpha)={2\over\alpha}\int_{-1}^{1}dt\,\log\bigg(\frac{\alpha t+\operatorname{arctanh}(t)}
{\operatorname{arctanh}(t)}\bigg)\,.
\end{align}
The substitution $t=\tanh v$ shows that $i_2(\alpha)$ agrees with the function defined in Ref.~\cite{Artemev:2025cev}. In the equal-mass limit,
\begin{equation}
 V_{++}(0)\Big|_{\beta=0}
=\frac12\frac{d}{d\alpha}[\alpha i_2(\alpha)].
\end{equation}

To determine $V'_{++}(0)$, we integrate \re{q-eq1} over $t$ and use the identity $\fint_{-1}^{1}dt/(t-t')=-\pi x(t')$. This gives
\begin{equation}
\int_{-1}^{1}dt\,[q_+(t)-1]=-\frac{2i\beta}{\pi}\int_{-1}^{1} dt\,x(t)\rchi(x(t))q_+(t) \,.
\end{equation}
Combining this identity with \re{Delta-QCD-app} and \re{app:V}, we obtain
\begin{align}\label{V''}\notag
{}&\Delta_{\rm QCD}=2-\alpha V_{++}(0) +\frac{\beta^2}{\pi^2}V'_{++}(0)\,,
\\[1.2mm]
{}&V'_{++}(0)=\frac{\pi^2}{\beta^2}\int_{-1}^{1}dt\, [q_+(t)-1] \,. 
\end{align}
  
\paragraph{Constant $\Delta_{\rm QCD}$.} For equal quark masses, $\beta=0$, the first relation in \re{V''} gives 
\begin{align}\label{Delta-QCD}
\Delta_{\rm QCD}\Big|_{\beta=0}=2-\alpha V_{++}(0)\Big|_{\beta=0}
=2-\frac{\alpha}{2}\frac{d}{d\alpha}[\alpha i_2(\alpha)].
\end{align}
For $\beta\ne0$, we start from \re{Delta-QCD-app}. Using the QCD symbol identity, we obtain from \re{app:Delta} , for $t\notin[-1,1]$,
\begin{align}
\Delta_{\rm QCD}
={}&t\int_{-1}^{1}\frac{q_+(t')}{t-t'}\,dt'
-\frac{2\alpha t}{\pi}\int_{-1}^{1}
\frac{\rchi(x(t'))q_+(t')}{t-t'}\,dt' -\int_{-1}^{1}\frac{x(t')\rchi(x(t'))q_+(t')}{t-t'}\,dt'.
\end{align}
To evaluate the first integral, we substitute \re{q-eq1} and interchange the integrations with the principal-value prescription. The inner integral is
\begin{align}
\fint_{-1}^{1}\frac{dt'}{(t-t')(t'-t'')}
=\frac{1}{t-t''}\left[\log\frac{t+1}{t-1}-\pi x(t'')\right].
\end{align}
Using the relation \re{V-Cauchy-integral}, we find
\begin{align}
\int_{-1}^{1}\frac{q_+(t')}{t-t'}\,dt' ={}&\tilde q_+(t)\log\frac{t+1}{t-1} -\frac{2i\beta}{\pi}\int_{-1}^{1}\frac{x(t')\rchi(x(t'))q_+(t')}{t-t'}\,dt'.
\end{align}
Combining these relations and applying \re{V-Cauchy-integral} once more gives
\begin{align}
\left(\frac{2i\beta}{\pi}+\frac1t\right)& \int_{-1}^{1}\frac{x(t')\rchi(x(t'))q_+(t')}{t-t'}\,dt'
 =\tilde q_+(t)\log\frac{t+1}{t-1}
-\frac{\alpha\pi}{i\beta}[\tilde q_+(t)-1]
-\frac{\Delta_{\rm QCD}}{t}.
\end{align}
At $t=i\pi/(2\beta)$, the coefficient of the integral vanishes, while the integral itself is regular. For $\beta>0$, define $\xi_1=\arctan[\pi/(2\beta)]$ and use $\log[(t+1)/(t-1)]=i(2\xi_1-\pi)$.
We obtain
\begin{align}\label{Delta1}
\Delta_{\rm QCD}=\frac{\pi^2}{2\beta^2}
\left[\alpha-
\left(\alpha-\beta+\frac{2\beta }{\pi}\xi_1\right)
\tilde q_+ \left(\frac{i\pi}{2\beta}\right)\right].
\end{align}
The value of $\tilde q_+$ follows from \re{q-int}. Pairing the contributions at $t'$ and $-t'$ yields
 \begin{align}
\tilde q_+ \left(\frac{i\pi}{2\beta}\right)
 =\exp\left[\frac{1}{2\beta}\int_0^1\frac{dt'}{{t'}^2+\pi^2/(4\beta^2)}\log\frac{\alpha+\beta+\operatorname{arctanh}(t')/t'}{\alpha-\beta+\operatorname{arctanh}(t')/t'} \right].
\end{align}
To compare with Ref.~\cite{Artemev:2025cev}, use its definitions $g_{\pm1}=G(\mp2i\xi_1/\pi)$, where $G$ is the ratio of scalar factors defined in Eq.~(3.9) of that reference. Its integral representation gives $\tilde q_+(i\pi/(2\beta))=g_{-1}$. The product identity (3.23) of that reference then yields  
\begin{equation}
\tilde q_+\left(\frac{i\pi}{2\beta}\right)
=\frac{\alpha+\beta+2\beta\xi_1/\pi}
{(\alpha-\beta+2\beta\xi_1/\pi)g_1}.
\end{equation}
Substitution into \re{Delta1} gives
\begin{align} 
\Delta_{\rm QCD}
&=\frac{\pi^2}{2\beta^2}
\left[\alpha-
\frac{\alpha+\beta+2\beta\xi_1/\pi}{g_1}\right].
\end{align}
This is precisely the first relation in Eq.~(6.1) of Ref.~\cite{Artemev:2025cev}, $\Delta_{\rm QCD}=G_+^{(1)}-G_-^{(1)}$. Together with \re{V''}, this determines the remaining constant $V'_{++}(0)$.

Finally, eliminating $V'_{++}(0)$ with \re{V''} and using the explicit expression for $V_{++}(0)$, we obtain the subtraction constants in the regularization scheme of Ref.~\cite{Artemev:2025cev}   
\begin{align}\label{d-QCD}
d_\pm
&=1\mp\frac{\Delta_{\rm QCD}}{2}
+\frac{(\alpha+\beta)^2i_2(\alpha+\beta)
-(\alpha-\beta)^2i_2(\alpha-\beta)}{8\beta}\,.
\end{align}  
In particular, $d_--d_+=\Delta_{\rm QCD}$, as required by \re{second1}. With these normalizations, the localization identities \re{localization-Dminus} and \re{localization-Dplus} coincide with Eqs.~(5.19) and (5.22) in \cite{Artemev:2025cev}.

\paragraph{Proof of the identity \re{reciprocal}.}

The first relation in \re{mod-com} implies
\begin{align}\label{comm0}
[\bm t,\bm\rho^{-1}]
&=\bm\rho^{-1}[\bm\rho,\bm t]\bm\rho^{-1}=\frac{\pi\lambda}{2}
\bm\rho^{-1}\left(|e_-\rangle\langle e_+|-|e_+\rangle\langle e_-|\right)\bm\rho^{-1}
+\frac{i\beta}{\pi}\bm\rho^{-1}|e_+\rangle\langle e_+|\bm\rho^{-1}\,.
\end{align}
We evaluate its kernel at $x=y=0$. The right-hand side involves the matrix elements $\langle0|\bm\rho^{-1}|e_\pm\rangle$ and $\langle e_\pm|\bm\rho^{-1}|0\rangle=\langle 0|(\bm\rho^t)^{-1}| e_\pm\rangle$, where $\bm\rho^t$ denotes the transposed operator \re{rho}. The kernels of $\bm\rchi^{-1}$ and $\bm K$ are symmetric, whereas $H(y-x)=-H(x-y)$. Hence $\bm\rho^{t}=\bm P\bm\rho\bm P$, where $\bm P$ is the parity operator, and the same holds for $\bm\rho^{-1}$. 
Since $e_+(0)=1$ and $\bm Pe_\pm=\pm e_\pm$, we find $\langle0|\bm\rho^{-1}|e_\pm\rangle=\rchi(0)Q_\pm(0)$ and $\langle e_\pm|\bm\rho^{-1}|0\rangle=\pm\rchi(0)Q_\pm(0)$. The relation \re{comm0} then yields
\begin{align}\label{comm1}
\langle0|[\bm t,\bm\rho^{-1}]|0\rangle
=\rchi^2(0)\left[
\pi\lambda Q_+(0)Q_-(0)+\frac{i\beta}{\pi}Q_+^2(0)\right]
=\pi\lambda\rchi^2(0)\,
\mathcal Q_+(0|\lambda)\mathcal Q_-(0|\lambda)\,,
\end{align}
where the second equality follows from \re{Q-odd-combination}.

The commutator on the left-hand side of \re{comm1} has the kernel $[t(x)-t(y)]\langle x|\bm\rho^{-1}|y\rangle$, so only the singular part of $\langle x|\bm\rho^{-1}|y\rangle$ at $x\to y$ contributes. Since $H(x-y)=2/[\pi(x-y)]+O(x-y)$ and $K(x-y)$ is smooth,
\begin{align}
\langle x|\bm\rho|y\rangle
={1\over\rchi(x)}\delta(x-y)
-\frac{2i\beta}{\pi^2}\operatorname{PV}\frac1{x-y}
+\text{regular terms}\,,
\end{align}
where PV denotes the principal-value prescription. We look for the inverse in the form
\begin{align}\label{PB}
\langle x|\bm\rho^{-1}|y\rangle=c(x)\,\delta(x-y)+s(x)\operatorname{PV}\frac1{x-y}+\text{regular terms}\,.
\end{align}
The condition $\bm\rho\bm\rho^{-1}=1$ and the Poincar\'e--Bertrand formula $\operatorname{PV}\frac{1}{x-z}\circ\operatorname{PV}\frac{1}{z-y}=-\pi^2\delta(x-y)+\dots$, with $\circ$ denoting integration over the intermediate point $z$, fix the coefficient functions 
\begin{align}\label{singularity}
c(x)=\frac{\rchi(x)}{\Delta_+(x)\Delta_-(x)}\,,\qqqquad s(x)=\frac{2i\beta}{\pi^2}\frac{\rchi^2(x)}{\Delta_+(x)\Delta_-(x)}\,.
\end{align}
The factor $t(x)-t(y)$ annihilates the $\delta-$function term in \re{PB}, cancels the pole in the second term and suppresses the regular terms. Using $t'(0)=\pi/2$, we obtain
\begin{align}
\langle0|[\bm t,\bm\rho^{-1}]|0\rangle
=t'(0)\,s(0)=\frac{i\beta}{\pi}
\frac{\rchi^2(0)}{\Delta_+(0)\Delta_-(0)}\,.
\end{align}
Comparing with \re{comm1} and \re{Q00}, we arrive at the relation \re{reciprocal}.

\paragraph{Wronskian relation.}
The functions $\mathcal Q_\pm(z|\lambda)$ satisfy the same Baxter equation \re{Q-TQ} as the functions $Q_\pm(z|\lambda)$, since the coefficients in the linear combination
\re{Q-odd-combination} are independent of $z$. We define their Wronskian as
\begin{equation}\label{seed-Q-Wronskian-definition}
 W(z)\equiv
 \mathcal Q_+(z|\lambda)\mathcal Q_-(z+2i|\lambda)
 -\mathcal Q_-(z|\lambda)\mathcal Q_+(z+2i|\lambda)\,.
\end{equation}
It follows from the Baxter equation \re{Q-TQ} that the Wronskian satisfies
\begin{equation}\label{seed-Q-Wronskian-shift}
 \Delta_+(z+2i)W(z)=\Delta_-(z-2i)W(z-2i)\,.
\end{equation}
To solve this first-order difference equation, we use the function $\mathcal Q_+(z|0)=Q_+(z|0)$. Its direct
continuation $\widetilde Q_+(z|0)=\tilde q_+(t(z))$ (see \re{V-Cauchy-integral}) depends on $z$ only through
$t(z)=\tanh(\pi z/2)$. Since $t(z+2i)=t(z)$, it obeys
$\widetilde Q_+(z+2i|0)=\widetilde Q_+(z|0)$.
For $-2<\Im z<0$, the points $z$ and $z+2i$ lie in the lower
and upper halves of the fundamental strip, respectively.
Applying the dressing prescription \re{Q-Baxter-continuation}
to these two values yields
\begin{align}\label{per}
 \Delta_+(z+2i)\mathcal Q_+(z+2i|0)
 =\Delta_-(z)\mathcal Q_+(z|0)\,.
\end{align}
Elsewhere, this identity is understood by analytic continuation.
Combining \re{seed-Q-Wronskian-shift} with \re{per} shifted by
$z\to z-2i$, we obtain
\begin{align}\label{quot}
 \frac{\Delta_+(z+2i)W(z)}{\mathcal Q_+(z|0)}
 =\frac{\Delta_+(z)W(z-2i)}{\mathcal Q_+(z-2i|0)}\,.
\end{align}
This relation implies that $R(z)=\Delta_+(z+2i)W(z)/\mathcal Q_+(z|0)$ is $2i$-periodic. The residues picked up by analytic continuation cancel between the two terms in $W(z)$, while the dressing factor and the denominator cancel the remaining jump. Since the Cauchy solution \re{q-int} has no zeros, $R(z)$ has no finite singularities.

We now use the growth conditions to show that $R(z)$ is constant. For $\rchi(z)=O(|z|^{-1})$, the Cauchy representation implies that
$\mathcal Q_+(z|0)$ and its inverse grow at most as powers of $z$. Together with the assumed subexponential growth of the physical Baxter solutions, this means that $R(z)$ grows more slowly than $e^{\pi|\Re z|}$ in both directions, uniformly in the fundamental strip. A regular $2i$-periodic function can be expanded in modes $e^{n\pi z}$, with integer $n$. Every nonzero mode grows at least as fast as $e^{\pi|\Re z|}$  along the real axis. The growth condition therefore leaves only the constant mode.
 
We can determine the constant by putting $z=-i$ in \re{seed-Q-Wronskian-definition} and taking into account the relations
\re{Q-special-raw}, \re{Q-Baxter-continuation}, and \re{Q-odd-combination}. 
 The terms proportional to $\beta/\lambda$ cancel between the two
products in $W(-i)$, giving
\begin{align}
    W(-i)=\frac{2i}{\Delta_+(i)\Delta_-(-i)}
     = \frac{2i}{\Delta_+(i)}\mathcal Q_+(-i|0)\,.
\end{align}
Thus the periodic quotient
\re{quot} equals $2i$ leading to
\begin{equation}\label{seed-Q-Wronskian}
W(z)=\mathcal Q_+(z|\lambda)\mathcal Q_-(z+2i|\lambda)
 -\mathcal Q_-(z|\lambda)\mathcal Q_+(z+2i|\lambda)=\frac{2i\,\mathcal Q_+(z|0)}{\Delta_+(z+2i)}\,,
\end{equation}
where the dressing factor $\Delta_+(z)$ is defined in \re{Q-dressing-factors}.

For $\beta=0$, the dressing factors and $\mathcal Q_+(z|0)$
are unity, leading to $W(z)=2i$.
Evaluating the Baxter equation and the Wronskian at $z=0$, and
using $\mathcal Q_+(-z)=\mathcal Q_+(z)$ and
$\mathcal Q_-(-z)=-\mathcal Q_-(z)$, gives
\begin{equation}\label{seed-origin-adjacent}
 \mathcal Q_+(2i|\lambda)
 =[1-2\pi\lambda\rchi(0)]\mathcal Q_+(0|\lambda),\qqqquad
 \mathcal Q_-(2i|\lambda)=\frac{2i}{\mathcal Q_+(0|\lambda)}
\,.
\end{equation}
The values $\mathcal Q_\pm(2i|\lambda)$ are understood as boundary values from within
the fundamental strip. These relations hold for $\beta=0$ and apply, in particular,
to the Ising symbol.
 
\section{Matrix elements of the spectral operator}\label{app:C}

In this appendix, we derive the relations \re{1st} and \re{eq:finite-pole}. In what follows, the
spectral parameter $\lambda$ is suppressed unless needed explicitly. We introduce the auxiliary states
\begin{align}\label{complex-pole-sources1}
 \ket{e_{a,\pm}^R}=(t_a-\bm t)^{-1}\ket{e_\pm}\,,\qquad
 \bra{e_{a,\pm}^{\,L}}=\bra{e_\pm}(t_a-\bm t)^{-1}.
\end{align} 
where $t_a$ is a pole of the rational functions in \re{AB}, and $\bm t$ denotes multiplication by
$t(x)=\tanh(\pi x/2)$.  For complex $t_a$, the states $ \ket{e_{a,\pm}^R}$
are not the Hermitian adjoints of $\bra{e_{a,\pm}^{\,L}}$.

The contribution of the pole $t_a$ to the logarithmic derivative of the spectral determinant $\partial_\lambda \log D(\lambda)$ is given by
\begin{align}\label{fa-app}
f_a(\lambda)={}&\left(\frac{\pi d_a}{2}-\frac{\pi^2\lambda c_a}{2}\right)
\bra{e_{a,+}^{\,L}}\bm\rho^{-1}\ket{e_{a,+}^R}+\frac{\pi^2\lambda c_a}{8}
\bra{e_{a,+}^{\,L}}\bm H\bm\rho^{-1}\bm H\ket{e_{a,+}^R}.
\end{align} 
We show below that both matrix elements can be expressed in terms of Baxter $Q$-functions
at $x_a$ and $x_a-2i$. These points satisfy $t(x_a)=t(x_a-2i)=t_a$ and $0<\Im x_a<2$.

Let $z_\star$ denote either $x_a$ or $x_a-2i$. At these points, the kernels \re{HK} take the form
\begin{align}
H(z_\star-y)=e_+(z_\star)e_a^L(y) \,,\qquad K(z_\star-y)={\pi\over 2}(z_\star-y)e_+(z_\star)e_a^L(y)\,,
\end{align}
where $e_+(x)$ is the reference state \re{epm} and $e_a^L(y)=\vev{e_{a,+}^{L}|y}=e_+(y)/(t_a-t(y))$. Substituting these identities into \re{Q-direct-continuation} gives
\begin{align}\label{tildeQ-star}
 \widetilde Q_\sigma(z_\star)
 =E_\sigma(z_\star)
 +\left(\frac{\pi\lambda z_\star}{2}+\frac{i\beta}{\pi}\right)
 \bra{e_{a,+}^{\,L}}\bm\rho^{-1}\ket{e_\sigma}
 -\frac{\pi\lambda}{2}
 \bra{e_{a,-}^{\,L}}\bm\rho^{-1}\ket{e_\sigma}.
\end{align}
Here $\sigma=\pm$, with $E_+(z_\star)=1$ and $E_-(z_\star)=z_\star$.
Taking the difference between the values at $x_a$ and $x_a-2i$,
and passing to the functions $\widetilde{\mathcal Q}_+=\widetilde Q_+$ and
$\widetilde{\mathcal Q}_-=\widetilde Q_-+
 i\beta\widetilde Q_+/(\pi^2\lambda)$, yields
\begin{align}\notag\label{sys-app}
 \widetilde{\mathcal Q}_+(x_a)-\widetilde{\mathcal Q}_+(x_a-2i)
 &=i\pi\lambda\bra{e_{a,+}^{\,L}}\bm\rho^{-1}\ket{e_+},\\
 \widetilde{\mathcal Q}_-(x_a)-\widetilde{\mathcal Q}_-(x_a-2i)
 &=2i+i\pi\lambda\bra{e_{a,+}^{\,L}}\bm\rho^{-1}\ket{e_-}
 -\frac\beta\pi\bra{e_{a,+}^{\,L}}\bm\rho^{-1}\ket{e_+}.
\end{align} 
 These relations allow us to express the matrix elements $\bra{e_{a,+}^{\,L}}\bm\rho^{-1}\ket{e_\pm}$ through the values 
$\widetilde{\mathcal Q}_\pm(z_\star)$. The latter are related to the Baxter
functions by
\begin{align}\notag 
& \widetilde{\mathcal Q}_\sigma(x_a|\lambda)
 =\Delta_+(x_a)\mathcal Q_\sigma(x_a|\lambda),
 \\[2mm]
& \widetilde{\mathcal Q}_\sigma(x_a-2i|\lambda)
 =\Delta_-(x_a-2i)\mathcal Q_\sigma(x_a-2i|\lambda),
 \qquad \sigma=\pm.
\end{align} 

At the next step, we establish the relation between the matrix elements in \re{fa-app} and  $\bra{e_{a,+}^{\,L}}\bm\rho^{-1}\ket{e_\pm}$. 
 We start with the identity  
\begin{align}\notag\label{der}
\partial_{t_a}\bra{e_{a,+}^{\,L}}\bm\rho^{-1}\ket{e_\sigma}{}&=-\bra{e_+}(t_a-\bm t)^{-2}\bm\rho^{-1}\ket{e_\sigma}
\\[1.2mm]\notag
{}&=-\bra{e_{a,+}^{\,L}}\bm\rho^{-1}\ket{e_{a,\sigma}^R}-\bra{e_+}(t_a-\bm t)^{-1}[(t_a-\bm t)^{-1},\bm\rho^{-1}]\ket{e_\sigma}
\\[1.2mm]
{}&=-\bra{e_{a,+}^{\,L}}\bm\rho^{-1}\ket{e_{a,\sigma}^R}   
 -\bra{e^L_{a,+}}\bm\rho^{-1}(t_a-\bm t)^{-1}
 [\bm\rho,\bm t](t_a-\bm t)^{-1}\bm\rho^{-1}\ket{e_\sigma}
\end{align}
The commutator in the last line has the finite-rank form (see \re{mod-com})
\begin{align}
(t_a-\bm t)^{-1}[\bm\rho,\bm t](t_a-\bm t)^{-1}
 ={}&\frac{\pi\lambda}{2}
 (\ket{e_{a,-}^R}\bra{e_{a,+}^{\,L}}-\ket{e_{a,+}^R}\bra{e_{a,-}^{\,L}})
 +\frac{i\beta}{\pi}\ket{e_{a,+}^R}\bra{e_{a,+}^{\,L}}\,.
\end{align}
Applying \re{der} for $\sigma=+$ and $\sigma=-$, and eliminating
the mixed matrix elements, gives
\begin{align}\notag\label{vev1-app}
&\bra{e_{a,+}^{\,L}}\bm\rho^{-1}\ket{e_{a,+}^R}=-\frac{1}{\widetilde{\mathcal Q}_+(x_a|0)}\bigg[
\left(1+\frac{\pi\lambda}{2}
\bra{e_{a,+}^{\,L}}\bm\rho^{-1}\ket{e_-}\right)
\partial_{t_a}\bra{e_{a,+}^{\,L}}\bm\rho^{-1}\ket{e_+}\\
&\hspace{30mm}-\frac{\pi\lambda}{2}
\bra{e_{a,+}^{\,L}}\bm\rho^{-1}\ket{e_+}
\partial_{t_a}\bra{e_{a,+}^{\,L}}\bm\rho^{-1}\ket{e_-}
\bigg].
\end{align} 
The denominator is the determinant of this linear system. Its expression in terms of the $\widetilde{\mathcal Q}_\pm$ values is
\begin{align}\notag
\widetilde{\mathcal Q}_+(x_a|0){}&=\frac{1}{2i}\big[
\widetilde{\mathcal Q}_+(x_a-2i)\widetilde{\mathcal Q}_-(x_a)
-\widetilde{\mathcal Q}_-(x_a-2i)\widetilde{\mathcal Q}_+(x_a)\big]
\\\notag
{}&=\left(1-\frac{\pi\lambda}{2}
\bra{e_{a,-}^{\,L}}\bm\rho^{-1}\ket{e_+}\right)
\left(1+\frac{\pi\lambda}{2}
\bra{e_{a,+}^{\,L}}\bm\rho^{-1}\ket{e_-}\right)\\
&\quad+\frac{\pi^2\lambda^2}{4}
\bra{e_{a,+}^{\,L}}\bm\rho^{-1}\ket{e_+}
\bra{e_{a,-}^{\,L}}\bm\rho^{-1}\ket{e_-}
+\frac{i\beta}{\pi}\bra{e_{a,+}^{\,L}}\bm\rho^{-1}\ket{e_+}
\end{align}
The first equality is the  Wronskian relation
\re{seed-Q-Wronskian}, expressed in terms of the dressed functions
above; the second follows by substituting \re{tildeQ-star}.

Substituting \re{sys-app} into \re{vev1-app} combines the source derivatives into a differential Wronskian. With the convention $\mathcal W(f,g)=fg'-gf'$, where the prime denotes differentiation with respect to $x_a$, we obtain
\begin{align}
 \bra{e_{a,+}^{\,L}}\bm\rho^{-1}\ket{e_{a,+}^R}
 =-\frac{\mathcal W\Big(
 \widetilde{\mathcal Q}_+(x_a)-\widetilde{\mathcal Q}_+(x_a-2i),
 \widetilde{\mathcal Q}_-(x_a)-\widetilde{\mathcal Q}_-(x_a-2i)\Big)}
 {\pi^2\lambda(1-t_a^2)\widetilde{\mathcal Q}_+(x_a|0)},
\end{align}
where the functions with shifted arguments are differentiated
with respect to the same variable $x_a$. The factor of $(1-t_a^2)$ in the denominator comes from 
 the change of variables  
$\partial_{t_a}=\frac{2}{\pi(1-t_a^2)}\partial_{x_a}$.

To evaluate the second matrix element in \re{fa-app}, we first use
\begin{align}\notag\label{H-app}
 \bm H\ket{e_{a,+}^R}
 &=2\big[\ket{e_{a,-}^R}-(x_a-i)\ket{e_{a,+}^R}\big],
 \\[2mm]
 \bra{e_{a,+}^{\,L}}\bm H
 &=-2\big[\bra{e_{a,-}^{\,L}}-(x_a-i)\bra{e_{a,+}^{\,L}}\big],
\end{align}
The first identity follows by commuting $(t_a-\bm t)^{-1}$ through
$\bm H$, using \re{H-com} and \re{c-rel}, and evaluating
$\int_{-1}^1dt/(t_a-t)=\pi(x_a-i)$ on the branch
$0<\Im x_a<2$. Transposition and $\bm H^t=-\bm H$ give the
second identity, including its minus sign. Consequently,
\begin{align}
\bra{e_{a,+}^{\,L}}\bm H\bm\rho^{-1}\bm H\ket{e_{a,+}^R}=-4\left[\bra{e_{a,-}^{\,L}}-(x_a-i)\bra{e_{a,+}^{\,L}}\right]
\bm\rho^{-1}
\left[\ket{e_{a,-}^R}-(x_a-i)\ket{e_{a,+}^R}\right].
\end{align}
The right-hand side involves the four matrix elements
$\bra{e_{a,\sigma}^{\,L}}\bm\rho^{-1}\ket{e_{a,\sigma'}^R}$,
with $\sigma,\sigma'=\pm$. For $\sigma=\sigma'=+$ the matrix element is given by
\re{vev1-app}. The remaining matrix elements can be found in a similar manner.
Namely, proceeding as in \re{der}, we evaluate $\partial_{t_a}\bra{e_{a,-}^{\,L}}\bm\rho^{-1}\ket{e_\sigma}$ and then use it to generalize \re{vev1-app} to find $\bra{e_{a,\sigma}^{\,L}}\bm\rho^{-1}\ket{e_{a,\sigma'}^R}$. As in the previous case, these matrix elements can be expressed through the boundary value $\widetilde{\mathcal Q}_+(x_a|0)$ and the differential Wronskians of the functions $\widetilde Q_\pm(x_a)$ and  $\widetilde Q_\pm(x_a-2i)$. In this way, we arrive at 
\begin{align}\label{complex-second-contraction}\notag
 &\bra{e_{a,+}^{\,L}}\bm H\bm\rho^{-1}\bm H\ket{e_{a,+}^R}=\frac{4}{\pi^2\lambda(1-t_a^2)\widetilde{\mathcal Q}_+(x_a|0)}
 \bigg\{4\widetilde{\mathcal Q}_+(x_a|0)
 +\frac{4i\beta}{\pi^2\lambda}\widetilde{\mathcal Q}_+'(x_a|0)
 \\\notag
 &\quad-\mathcal W\Big(
 \widetilde{\mathcal Q}_+(x_a)+\widetilde{\mathcal Q}_+(x_a-2i),
 \widetilde{\mathcal Q}_-(x_a)+\widetilde{\mathcal Q}_-(x_a-2i)\Big)\\
 &\quad+\frac{2\beta}{\pi^2\lambda}\Big[
 \mathcal W\big(\widetilde{\mathcal Q}_+(x_a),\widetilde{\mathcal Q}_-(x_a)\big)
 -\mathcal W\big(\widetilde{\mathcal Q}_+(x_a-2i),\widetilde{\mathcal Q}_-(x_a-2i)\big)
 \Big]\bigg\}.
\end{align}
Finally, substituting \re{1st} and \re{complex-second-contraction}
into \re{fa-app} gives the finite-pole contribution \re{eq:finite-pole}.
 
\bibliographystyle{JHEP}    

\bibliography{QCD2}
\end{document}